\documentclass[12pt]{article}

\usepackage{vmargin}
\usepackage{amsmath}
\usepackage{amssymb}
\usepackage[dvips]{graphicx}
\usepackage{rotating}
\usepackage{psfrag}
\usepackage{fancyhdr}
\usepackage{mathrsfs}
\usepackage{color}

\setmarginsrb{3cm}{2cm}{3cm}{2cm}{1cm}{1cm}{1cm}{1cm}

\newcommand{\be}{\begin{equation}}
\newcommand{\ee}{\end{equation}}
\newcommand{\ba}{\begin{eqnarray}}
\newcommand{\ea}{\end{eqnarray}}

\newcommand{\nn}{\nonumber\\}
\newcommand{\la}{\langle}
\newcommand{\ra}{\rangle}
\newcommand{\e}{&\!\!=\!\!&}
\newcommand{\eq}{&\!\!\equiv\!\!&}
\newcommand{\m}[1]{$\mathop{#1}$}

\author{E. Joung\footnote{joung@apc.univ-paris7.fr}\ ,
J. Mourad\footnote{mourad@apc.univ-paris7.fr} \\
APC\footnote{Unit\'e mixte du CNRS UMR 7164}\ , Universit\'e  Paris 7 \\
B\^atiment Condorcet, F-75205 Paris Cedex 13, France \\
{}\\
K. Noui\footnote{noui@lmpt.univ-tours.fr} \\
LMPT\footnote{Unit\'e mixte du CNRS UMR 6083}\ , Universit\'e  de Tours, F\'ed\'eration Denis Poisson \\
 Parc de Grandmont, 37200 Tours \\}

\title{\bf Three Dimensional Quantum Geometry \\
and Deformed Poincar\'e Symmetry}

\date{}

\begin{document}

\sloppy

\maketitle

\begin{abstract}
We study a three dimensional non-commutative space emerging in the
context of three dimensional Euclidean quantum gravity. Our
starting point is the assumption that the isometry group is
deformed to  the Drinfeld double $\mathcal{D}(SU(2))$\,. We
generalize to the deformed case the construction of $\mathbb{E}^3$
as the quotient of its isometry group $ISU(2)$ by $SU(2)$\,. 
We show that the algebra of functions on $\mathbb{E}^3$ becomes the
non-commutative algebra of $SU(2)$ distributions, $C(SU(2))^*$\,,
endowed with the convolution product. This construction gives the
action of $ISU(2)$ on the algebra and allows the determination of
plane waves and coordinate functions. In particular, we show that:
$(i)$ plane waves have bounded momenta; $(ii)$ to a given momentum
are associated several $SU(2)$ elements leading to an effective
description of an element in $C(SU(2))^*$ in terms of several physical
scalar fields on $\mathbb E^3$; $(iii)$ their product leads to a
deformed addition rule of momenta consistent with the bound on the
spectrum. We generalize to the non-commutative setting the
\emph{local} action for a scalar field. Finally, we obtain, using
harmonic analysis, another useful description of the algebra as
the direct sum of the algebra of matrices. The algebra of matrices
inherits the action of  $ISU(2)$\,: rotations leave the order of the
matrices invariant whereas translations change the order in a way
we explicitly determine.

\end{abstract}

\newpage

\tableofcontents

\newpage

\section{Introduction}

It is commonly believed that space-time cannot be described
in terms of usual differential geometry at arbitrary small scales.
A complete and consistent theory of quantum gravity
that would give a precise description of space-time at the Planck scale
is however still missing.
Many models have been proposed in the literature
where space-time appears non-commutative, fuzzy or
discrete \cite{fuzzy,discrete}.
In string theory for example \cite{sw},
the low energy effective theory on D-branes with an external B-field
is described in terms of quantum field theory (QFT) on the
 non-commutative Moyal space.
In loop quantum gravity \cite{discrete}, space-time is argued to
be discrete because geometrical operators such as the area and the
volume operators  have discrete spectra on the spin-networks
states space. These two approaches are very different at the
technical as well as the conceptual levels but raise the same
fundamental question of the Poincar\'e invariance at the Planck
scale. Is it broken, hidden or deformed? Whatever the answer is,
one can consider the  construction of a QFT on such spaces keeping
in mind that  invariance under the Poincar\'e group is crucial for
standard QFT. In this respect, a twisted version of the Poincar\'e
group was conjectured to replace the Poincar\'e group for the
non-commutative field theory on the Moyal plane \cite{chaichian}.
The consequences of the invariance under the twisted Poincar\'e
group were studied \cite{antichaichian} where it was shown that
the twisting in not observable in the S-matrix.

The same issue of Poincar\'e invariance has also been discussed in
the loop quantum gravity (LQG) and spin-foam (SF) frameworks. LQG
and SF models offer respectively a canonical and covariant (or
path integral) quantization of gravity.
In particular, it has been shown, in the
four dimensional case, that states of quantum geometries which are unphysical 
are eigenstates of
area operators with discrete eigenvalues. This result has
motivated the idea that Poincar\'e invariance must be broken at
the Planck scale and a phenomenology of Poincar\'e invariant
broken physics has emerged in the literature (see \cite{DSR} for
instance). However, it is worth mentioning that there is, up to
our knowledge, no clear relationship between discrete quantum
geometry emerging in the context of LQG and an eventual Poincar\'e
symmetry breaking.

In the three dimensional case, the situation is somehow simpler
\cite{FL,Noui}. First, it was argued that a self-gravitating
scalar Euclidean QFT (with no cosmological constant) is
effectively described by a non-commutative QFT with no-gravity
\cite{FL}. Newton's constant $G$ or equivalently the Planck length
${\ell_P}=G\hbar$ encodes the non-commutativity of the spacetime:
the algebra of functions on the Euclidean space-time is
non-commutative and its product, denoted $\star$\,, is a deformation
of the standard point-wise product with deformation parameter
$\ell_P$\,. Therefore, one explicitly sees, in such models, how
quantum gravity effects could be encoded in a non-commutative
space-time. The Drinfeld double $\mathcal{D}(SU(2))$ of the
classical Lie group $SU(2)$ plays the role of the \emph{isometry
algebra} of the non-commutative space and is a deformation of the
group algebra of the Euclidean group $ISU(2)$ \cite{Bais} which is
the isometry group of the classical Euclidean space.
 Therefore, QFT is no longer
covariant under the usual Euclidean  group but  is covariant under
a deformed version of it. More precisely the action of any
symmetry element $\xi$ (in the quantum double) on a field (viewed
as a representation of the quantum double) is not modified, but
its action on a product of two fields $\phi_1$ and $\phi_2$ is
modified compared to the classical case and is such that 
\be
	\xi\rhd(\phi_1\star \phi_2)=\sum_{(\xi)}(\xi_{(1)}\rhd\phi_1) \star (\xi_{(2)}\rhd\phi_2)\,,
\ee
where $\Delta(\xi)=\sum_{(\xi)}\xi_{(1)}\otimes \xi_{(2)}$
is the $\mathcal{D}(SU(2))$ co-product.
This co-product is not co-commutative.
This was the starting point for the construction
of a scalar self-gravitating quantum field theory \cite{Noui}.

In fact, as usual in non-commutative geometry,
the non-commutative space is indirectly defined
through its non-commutative algebra of functions.
In the case we are studying, the momentum space is no longer a vector space,
it is a curved manifold closely related to the Lie group $SU(2)$\,.
More precisely, to each $SU(2)$ element $u$
one associates a three dimensional momentum $\vec P(u)$\,;
the sum $\vec{P}_{tot}$ of two momenta $\vec{P}(u)$ and $\vec{P}(v)$
is obtained from the $SU(2)$ group structure as follows:
\be
    \vec P_{\text{tot}}=\vec P(uv)\,.
\ee
The momentum space is then curved and has the topology of the
three dimensional sphere. The compactness of momentum space
 has the important effect of eliminating
 the potential ultra-violet divergences in QFT.
The price to be paid is a deformation of the addition rule of
momenta which is no more commutative. In a more formal language,
the co-product of the isometry algebra is deformed compared to the
standard case and becomes non co-commutative.

The aim of this article is to  study the non-commutative geometry
associated to the quantum double. We identify the relevant
non-commutative algebra replacing the algebra of functions on the
manifold $\mathbb{E}^3$ and we exhibit its realizations in
Euclidean space with a star product. Moreover, we show that the
non-commutative space has a discrete structure: the coordinate functions
satisfy the $\mathfrak{su}(2)$ lie algebra and have a discrete spectrum. 
The discreteness is physically
expected from the compactness of the momentum space.

Our starting point is the deformed isometry group of the manifold, the
Euclidean group $ISU(2)$ is replaced by its deformed version the
Drinfeld double ${\cal{D}}(SU(2))$\,. The latter contains $ISU(2)$ but has
a twisted co-algebra structure. We then generalize the construction
of $\mathbb{E}^3$ as the quotient of its isometry group $ISU(2)$
by $SU(2)$ to the deformed case. We show that the resulting
algebra is the non-commutative algebra of $SU(2)$ distributions
endowed with the convolution product, denoted $C(SU(2))^*$\,. 
This construction gives the action of $ISU(2)$ on the algebra and
consequently allows the determination of plane waves and
coordinate functions. Then we show that surprisingly  $C(SU(2))^*$ cannot be 
represented as a deformation
of the  commutative algebra of functions on $\mathbb E^3$.
This is due to the fact that the previously mentionned map $\vec P(u)$ cannot be
injective
 or in more formal terms
$\mathbb E^3$ and $SU(2)$ are not homeomorphic. However, we exhibit an
(family of) injective mapping(s) from $C(SU(2))^*$ to the direct
sum of three sub-spaces of $C(\mathbb E^3)$\,, each of them having a
good physical interpretation. Furthermore, This mapping is
compatible with the action of the Euclidean group and gives rise
to a star product which can be used to construct a local action for a scalar
field.
 We then
consider another useful description of the algebra obtained using
harmonic analysis as the direct sum of the algebra of matrices.
This makes clear the discreteness of the non-commutative space.
The algebra of matrices inherits the action of the $ISU(2)$
elements: rotations leave the order of the matrices invariant
whereas translations change the order in a way we determine.

The plan of the paper is as follows. In Section 2, we recall the
definition of the quantum double $\mathcal{D}(SU(2))$\,. We
exhibit a family of morphisms between the group algebra $\mathbb
C[ISU(2)]$ and $\mathcal{D}(SU(2))$ which makes clear that
$\mathcal{D}(SU(2))$ is a deformation of the group algebra of the
Euclidean group $ISU(2)$\,. We show that there is an
ambiguity in the definition of the momenta in
$\mathcal{D}(SU(2))$\,. A very similar ambiguity is also present in
full Loop Quantum Gravity. In Section 3, we construct the
non-commutative algebra using the quantum double. The basic idea
mimics the classical situation: the classical space $\mathbb E^3$
can be defined as the homogeneous space $ISU(2)/SU(2)$\,. In the
deformed case, we find that what replaces the algebra of functions
on $\mathbb E^3$ is the  non-commutative algebra $C(SU(2))^*$ of
$SU(2)$ distributions endowed with the convolution product. In
Section 4, we exhibit a morphism between $C(SU(2))^*$ and the
space $C_{{\ell_P}}(\mathbb E^3)$ which, viewed as a vector space,
is the direct sum of three-subspaces of $C(\mathbb E^3)$\,:
two of them being the set of functions on $\mathbb E^3$ with a
spectrum strictly bounded by ${\ell_P}^{-1}$ and the last one the
set of functions on $\mathbb E^3$ whose spectrum belongs to the
sphere of radius ${\ell_P}^{-1}$. The algebra structure of
$C_{{\ell_P}}(\mathbb E^3)$ is obviously non-commutative and leads
to a $\star$-product. We finally write an action for a scalar
field in this non-commutative space, we postpone its detailed
study  for the future. In Section 5, we show that $C(SU(2))^*$
admits in fact a discrete structure. More precisely, we introduce
a Fourier map on $C(SU(2))^*$ whose image is the space of complex
matrices $\bigoplus_{n\in\mathbb{N}} \text {Mat}_{n\times
n}(\mathbb C)$\,. This makes explicit the fuzzy space formulation
of the non-commutative space which  appears as a collection of
concentric fuzzy spheres of different radii. Then, we determine
the action of the \emph{isometry algebra} $\mathcal{D}(SU(2))$ on
the fuzzy space: rotations leave each fuzzy sphere invariant
whereas translations define finite difference operators which send
points on a fuzzy sphere to points on neighboring spheres. In
Section 6, we study the correspondence between the continuous
formulation of $C(SU(2))^*$ and its discrete formulation in terms
of matrices. This relation gives some insight on the fuzzy
geometry and its classical limit. We conclude with some remarks on
the construction of a QFT on the non-commutative space. This QFT
is interesting for many reasons, the most important being the
possibility of eliminating UV divergences. In the Appendix we use
Schwinger's oscillator representation of  $SU(2)$ to
determine the  action of translations in terms of a couple of creation
and annihilation operators.

\section{The Quantum double $\mathcal{D}(SU(2))$:
a deformation of the Euclidean group $ISU(2)$}

The quantum double $\mathcal{D}(SU(2))$ is a deformation of the
group algebra of the Lie group $ISU(2)$ \cite{Bais}. This Hopf
$*$-algebra was extensively studied in the context of
combinatorial quantization of Chern-Simons theory \cite{Comb} with
the Euclidean group $ISU(2)$\,; such a theory is equivalent
\cite{Witten} (up to some discrepancies \cite{Matschull}) to three
dimensional Euclidean gravity without a cosmological constant. In
that context, the Newton constant $G$ or equivalently the Planck
length ${\ell_P}=\hbar\,G$ plays the role of the deformation
parameter.

\subsection{Definition of the quantum double $\mathcal{D}(SU(2))$}

The quantum double $\mathcal{D}(SU(2))$ is defined as a vector
space by the tensor product:
\be
    \mathcal{D}(SU(2))=C(SU(2))\otimes \mathbb C[SU(2)]\, ,
\ee where $C(SU(2))$ is a suitable set of complex functions on
$SU(2)$ and $\mathbb C[SU(2)]$ is the group algebra of $SU(2)$\,.
 When $u\in SU(2)\,$, and for elements of
$\mathcal{D}(SU(2))$ of the form, $(f\otimes u)$\,,  the  Hopf $*$-algebra structure is given by
\ba
    \label{DSU(2)_definition}
    &\text{Product} &
    \quad : \qquad
    (f_1\otimes u_1)\cdot(f_2\otimes u_2) =
    (f_1 \cdot \text{ad}_{u_1}f_2\otimes u_1u_2) \nn
    &\text{Co-product} &
    \quad : \qquad
    \Delta_{\mathcal{D}(SU(2))}(f\otimes u) = \sum_{(f)}\,(f_{(1)}\otimes u)
    \otimes(f_{(2)}\otimes u) \nn
    &\text{Unit} &
    \quad : \qquad
    (1\otimes e)  \nn
    &\text{Co-unit}&
    \quad : \qquad
    \varepsilon(f\otimes u)=f(e) \nn
    &\text{Antipode}&
    \quad : \qquad
    S(f\otimes u)=(\iota\,\text{ad}_{u^{-1}}f\otimes  u^{-1})\nn
    &\text{Complex conjugate}&
    \quad : \qquad
    (f\otimes u)^*=(\overline{\text{ad}_{u^{-1}}f}\otimes  u^{-1})\,,
\ea
where the adjoint map $\text{ad}$ and the inverse map $\iota$
are respectively defined by
\ba
    (\text{ad}_uf)(v)=f(u^{-1}vu) \,,\qquad
    (\iota f)(u) = f(u^{-1})\,,
\ea
and we used the Sweedler notation. The explicit
expression for the $C(SU(2))$ coproduct is
\be\label{C(SU(2)) coproduct}
    \Delta_{C(SU(2))}(f)(u\otimes v) =  \sum_{(f)} f_{(1)}(u)\, f_{(2)}(v) = f(uv)\,.
\ee
These relations are then extended by linearity, morphism (for
the product, co-product and co-unit) and anti-morphism (for the
antipode and complex conjugate operation) when $u$ is a general
element of $\mathbb C[SU(2)]\,$. Notice that $\mathbb C[SU(2)]$
and $C(SU(2))$ are two sub-Hopf $*$-algebras.

\subsection{Drinfeld double $\mathcal{D}(SU(2))$ as a deformation of $\mathbb{C}[ISU(2)]$}

In order to see how $\mathcal{D}(SU(2))$ can be viewed as a
deformation of $\mathbb{C}[ISU(2)]$\,, we start by recalling the
structure of the group algebra $\mathbb{C}[ISU(2)]$\,.

\subsubsection*{Group algebra $\mathbb{C}[ISU(2)]$}

As previously
for the quantum double, $\mathbb{C}[ISU(2)]$ is defined as a
vector space by the tensor product:
\be
    \mathbb{C}[ISU(2)]=\mathbb{C}[\mathbb R^3 \rtimes SU(2)]
    =\mathbb C[\mathbb R^3]\otimes \mathbb{C}[SU(2)]\,,
\ee
where $\mathbb R^3$ is the group of standard Euclidean
translations whose elements will be denoted $\mathcal{T}_x$ for $x \in \mathbb R^3$\,.
We define the group algebra $\mathbb C[\mathbb R^3]$
as the set of elements $\int d^3x\,\tilde f(x)\,\mathcal{T}_x$ with
 $\tilde f$ a function (or a distribution)  of compact support.
$\mathbb C[\mathbb R^3]$  can be canonically identified with a
sub-algebra  $C(\mathbb{R}^3)$ of the algebra of functions: to any
element $\int d^3x\,\tilde f(x)\,\mathcal{T}_x$\,, one associates the
function $f(p)=\int d^3x\,\tilde f(x)\,e^{-i\,\vec p\cdot\vec x}$ whose
Fourier transform is of compact support. In particular, to  the
translation $\mathcal{T}_x$ is associated the function $T_x(p)=e^{-i\,\vec p\cdot\vec x}$\,.
Finally, $\mathbb{C}[ISU(2)]$ can be identified with the  vector space:
\be
    \mathbb{C}[ISU(2)] \simeq C(\mathbb{R}^3)\otimes \mathbb C[SU(2)]\,.
\ee
The group algebra $\mathbb{C}[ISU(2)]$ is in fact a  Hopf
$*$-algebra. When $u\in SU(2)$ the Hopf $*$-algebra structure 
is given for elements of the form
$(f\otimes u)\in C(\mathbb{R}^3)\otimes \mathbb C[SU(2)]$\,, by
\ba
    &\text{Product} &
    \quad : \qquad
    (f_1\otimes u_1)\cdot(f_2\otimes u_2) = (f_1 \cdot \text{R}_{u_1}f_2\otimes u_1u_2) \nn
    &\text{Coproduct} &
    \quad : \qquad
    \Delta_{\mathbb{C}[ISU(2)]}(f\otimes u)
    = \sum_{(f)}\,(f_{(1)}\otimes u)\otimes(f_{(2)}\otimes u) \nn
    &\text{Unit} &
    \quad : \qquad
    (1\otimes e) \nn
    &\text{Counit}&
    \quad : \qquad
    \varepsilon(f\otimes u)=f(e) \nn
    &\text{Antipode}&
    \quad : \qquad
    S(f\otimes u)=(\iota\,\text{R}_{u^{-1}}f\otimes  u^{-1}) \nn
    &\text{Complex conjugate}&
    \quad : \qquad
    (f\otimes u)^*=(\overline{\text{R}_{u^{-1}}f}\otimes  u^{-1})\,,
\ea
where the inverse map $\iota$ is defined as for
${\cal D}(SU(2))$ and $\text{R}_u$ is given by 
\m{(\text{R}_uf)(p)\equiv f(R(u^{-1})\,p)}\,, $R(u)$ being the vectorial representation of
$SU(2)$\,. The  co-product on $C(\mathbb{R}^3)$  explicitly reads
\be\label{C(R^3) coproduct}
    \Delta_{C(\mathbb{R}^3)}(f)(p\otimes q)
    =\sum_{(f)} f_{(1)}(p)\, f_{(2)}(q) = f(p+q)\,.
\ee
Contrary to the ${\cal D}(SU(2))$ case, this  co-product is
co-commutative. These relations are extended by linearity,
morphism or anti-morphism to any element of $\mathbb{C}[SU(2)]$\,.

Notice that the set of  elements $\{\,(T_x\otimes u)\,\} \subset
\mathbb{C}[ISU(2)]$ where $T_x(p)=e^{-i\,\vec p\cdot\vec x}$ equipped with the
product inherited from the Hopf algebra  is isomorphic to the
Euclidean group.

\subsubsection*{Algebra morphisms between $\mathbb{C}[ISU(2)]$ and ${\cal D}(SU(2))$}

The algebras $\mathbb{C}[ISU(2)]$ and $\mathcal{D}(SU(2))$ are in
fact closely related. To clarify this point, we are going to
construct an algebra morphism $\varphi$ between these two
algebras. We require this morphism to be trivial on the 
$\mathbb C[SU(2)]$ part, namely 
\m{\varphi(f\otimes u)\equiv \varphi(f)\otimes u}\,; 
therefore $\varphi$ is viewed as a mapping from $C(\mathbb{R}^3)$ to $C(SU(2))$\,. 
From the algebra structures of $\mathbb C[ISU(2)]$ 
and $\mathcal{D}(SU(2))$\,, it is immediate
to see that $\varphi$ must satisfy the following property:
\be\label{cov req}
    \varphi(\text{R}_u f)= \text{ad}_u \varphi(f)\,,
\ee
for any $SU(2)$ element $u$\,.

In order to find the solutions $\varphi$ to this  equation, it
will be useful to introduce explicit parameterizations of $SU(2)$.
Actually, two different parameterizations will be useful. In the
first one, we identify $SU(2)$ with the three-sphere
\m{S^3=\{(\vec{y},y_4) \in \mathbb R^4 \;\vert\; {\vec
y\,}^2+y_{4}^2=1\}} and then the fundamental representation of 
any $u \in SU(2)$ is given by
\be\label{param y}
    u(\vec{y},y_4) = y_4 - i \,\vec{y} \cdot \vec{\sigma}\,,
\ee 
with $ \vec \sigma$ being the Pauli matrices. In the second
one, any group element $u$, but the identity $u=e=1$ and its antipode $u=e_A=-1$,
 is characterized by a rotation angle
(its conjugacy class) $\theta\in\,]0,2\pi[\,$\ together with a
unit vector $\vec n\in S^2$ as follows: 
\be\label{parametrisation of su2}
    u(\theta,\vec n) = \exp(-\,i\,\theta\, n^a\mathcal{J}_a)\,,
\ee 
where $\mathcal{J}_a$ are the generators of the Lie algebra
$\mathfrak{su}(2)$\,; they satisfy the relation
$[\mathcal{J}_a,\mathcal{J}_b]=i\,\epsilon_{abc}\,\mathcal{J}_c$\,.
Note that 
the \m{(\theta,\vec n)}-parametrization is not well-defined at $e$
and $e_A$ which are respectively
parameterized by $(\theta=0,\vec{n})$ and $(\theta=2\pi,\vec{n})$
for any unit vector $\vec{n}$\,. We are now ready to give
explicitly the solutions to eq. (\ref{cov req}) for $\varphi$\,: the map $\varphi$ is completely
characterized by an arbitrary function $\Pi$ defined on $[-1,1]$
according to: \be\label{morphi}
    \varphi(f)\big(u(\vec{y},y_4)\big)
    =f\big({\ell_P}^{-1}\,\Pi(y_4)\,\vec y\,)\,.
\ee
The Planck length ${\ell_P}$ has been introduced for dimensional purposes only
and from now on we  assume $\mathop{\Pi(1)=1}$\,.
Using the other parametrization, $\varphi$ reads
\be\label{morph}
    \varphi(f)\big(u(\theta,\vec n)\big)
    =f\big({{\ell_P}}^{-1}\,\Xi(\theta)\,\vec n\,\big)\,
\ee
with $\Xi(\theta)=\sin(\theta/2)\,\Pi\big(\cos(\theta/2)\big)$\,.

To understand some fundamental properties of $\varphi$\,,
it is particularly interesting to consider the images of the momentum coordinates functions
$\mathop{p_a\in C(\mathbb{R}^3)}$\, given by the
 $SU(2)$ functions $P_a$\,:
\ba\label{def de P}
    P_a(u) \equiv \varphi(p_a)(u) = {\ell_P}^{-1}\,\Pi(y_4)\,y_a
    ={\ell_P}^{-1}\,\Xi(\theta)\,n_a\,.
\ea
Note that the norm of the momentum vector $|\vec P|$
is function of the conjugacy class ($y_4$ or $\theta$) of $SU(2)$ 
and it vanishes on the identity $e$ and its antipode $e_A$\,.
Mathematically, there is of course no way to distinguish between
two different choices of $\Pi$ (or $\Xi$\,). Therefore, there exists an
ambiguity which is  very similar to those ambiguities that exist
in full Loop Quantum Gravity \cite{Ale}. However, the choice of a
morphism is physically rather important because it determines what
will be the momenta in our theory. Indeed, as we will see in the
sequel, $\vec P$ generate translations in the non-commutative space
whose isometry algebra is $\mathcal{D}(SU(2))$\,.

Another important property of $\varphi$ is that it is not invertible. In fact
 there is no
continuous and monotonous function $\Pi(y_4)$  which
vanishes on $y_4=\pm1$\,.
This is a signature of the fact that one cannot cover $S^3\simeq
SU(2)$ with a single coordinate patch.
This point will have important physical consequences in the sequel. We can
however 
 assume that $|\vec P|$ is invertible in a vicinity
$\mathcal V$ of the identity $e\in SU(2)$;
 then the restriction of $\varphi$ on $\mathcal V$ defines a bijection
from $\mathcal V$ to its image $\mathcal W \subset \mathbb{R}^3$. As a result,
given a vector $\vec{p} \in \mathcal W$,
we reconstruct an element $u(\vec y,y_4)$ of $SU(2)$ with:
\be\label{inverse of p}
    \vec y={\ell_P}\, \rho(p)\, \vec{p},\quad
    y_4={\sqrt{1-{\ell_P}^2\,p^2\,\rho^2(p)}}
   \,,
\ee
where $\rho$ is function of $p=\vert \vec p \vert$
related to $\Pi$ by the relation:
\be
    \rho(p)\,\Pi\left(\sqrt{1-{\ell_P}^2\,p^2\,\rho^2(p)}\right)= 1 \,.
\ee

\subsubsection*{The co-algebra structures: addition rule of momenta}

Contrary to the algebra structure, the co-algebra structures of
two Hopf $*$-algebras are obviously different. The co-product of the momenta
coordinates functions \m{(P_a\otimes e)\in\mathcal{D}(SU(2))}
can be expanded in powers of ${\ell_P}$ as follows:
\be\label{coproduct P}
    \Delta_{C(SU(2))}(P_a)=P_a\otimes 1 + 1 \otimes P_a +
    {\ell_P}\,\epsilon_{abc}\,P_b\otimes P_c + \mathcal{O}({{\ell_P}}^2)\,,
\ee
where $\epsilon_{abc}$ is the totally antisymmetric tensor with
$\epsilon_{123}=+1$\,. Comparing with the standard group-like
$C(\mathbb{R}^3)$ co-product,
\be
    \Delta_{C(\mathbb{R}^3)}(p_a)=p_a\otimes 1+1\otimes p_a\,,
\ee
it becomes clear that the co-algebra structure (\ref{coproduct P})
of $\mathcal{D}(SU(2))$ is a deformation (in the sense of
Drinfeld) of the co-algebra structure of $\mathbb{C}[ISU(2)]$ with
a deformation parameter given by ${\ell_P}$\,. According to Drinfeld
the two co-products are related by a twist operator which can be
explicitly found order by order in $\ell_P$\,.

The co-product is of course related to the addition rule of momenta. 
Given two momenta $\vec{p}$ and ${\vec{q}}$ in $\mathcal W$ 
their sum is given by
\be
    \Delta_{C(SU(2))}(\vec{P})\big(P^{-1}(\vec{p})\otimes P^{-1}(\vec{q})\big)
    =\vec{P}(P^{-1}(\vec{p})\,P^{-1}(\vec{q})) \,.
\ee It is important to note that this formula makes sense when
$P^{-1}(\vec{p})\,P^{-1}(\vec{q}) \in \mathcal V$\,. Using the
parametrization (\ref{param y}) and the inversion formula
(\ref{inverse of p}), one can write the previous sum more
explicitly as follows: 
\ba
  && \Pi\left(\sqrt{1-\left({\ell_P\,\rho(p)\,p}\right)^2}\sqrt{1-\left({\ell_P\,\rho(q)\,q}\right)^2}
  - \left({\ell_P}\right)^2 \rho(p)\,\rho(q)\,\vec{p}\cdot {\vec{q}}\right)\times\\
   && \;\times
    \left(\rho(q)\,{\vec{q}}\sqrt{1-\left({\ell_P\,\rho(p)\,p}\right)^2} 
    + \rho(p)\,{\vec{p}}\sqrt{1-\left({\ell_P\,\rho(q)\,q}\right)^2} +
    {\ell_P}\,\rho(p)\,\rho(q)\,\vec{p}\wedge {\vec{q}}\right).
    \nonumber
\ea

Let us illustrate the construction with explicit examples where the co-product
and then the addition rule of momenta have quite simple expressions.
The first example introduced in the literature consists in taking  $\Pi=1$\,.
It is clear that this choice makes the function $\varphi$ not invertible in the whole $SU(2)$
but we restrict it to a vicinity of the identity $\mathcal V$ where it is invertible.
The momenta coordinates functions are given by
$\mathop{\vec P(u)={\ell_P}^{-1}\,\vec y}$
(resp. $\vec{P}(u)={\ell_P}^{-1}\sin(\theta/2)\,\vec{n}$).
This choice leads to the following closed formula for the co-product of ${P_a}$\,:
\be\label{Pi=1}
    \Delta_{C(SU(2))}(P_a)=P_a\otimes P_4+
    P_4\otimes P_a+ {{\ell_P}}\,\epsilon_{abc}\,P_b\otimes
    P_c\,,
\ee
where $P_4$ is worth written in the $(\theta,\vec{n})$ parametrization:
$P_4={\ell_P}^{-1}\,\cos(\theta/2)$\,.
The addition rule of momenta follows immediately
\be
	\vec{q}\sqrt{1-(\ell_P p)^2}  +
    \vec{p} \sqrt{1-(\ell_P q)^2} + {\ell_P}\,\vec{p}\wedge\vec{q}\,,
\ee
when $P^{-1}(\vec{p})$\,, $P^{-1}(\vec{q})$ and 
$P^{-1}(\vec{p})\,P^{-1}(\vec{q})$ belong to $\mathcal V$\,.

Another possibility is to first consider $SO(3)$ instead of $SU(2)$
and then to choose $\Pi(y_4)=\text{sgn}(y_4)=y_4/\vert y_4 \vert$.
This choice has been made in the literature\cite{FM}
but leads to some difficulties related to the fact that
$\varphi$ becomes not only discontinuous at $y_4=0$
but above all not defined at these points.

\section{Convolution algebra $C(SU(2))^*$ from $\mathcal{D}(SU(2))$}

In the previous sections, we introduced all the ingredients
to construct the non-commutative space  with isometry algebra given by
the quantum double $\mathcal{D}(SU(2))$\,.
This non-commutative space is characterized as usual by its
 algebra of functions interpreted
as functions on the non-commutative space.
The basic idea is to define the non-commutative algebra of functions as the space of $\mathcal{D}(SU(2))$-linear
forms which are invariant under the action of ${\mathbb C}[SU(2)]$\,.

To illustrate this idea, let us recall how this works in the undeformed case:
the Euclidean manifold $\mathbb E^3$ admits $ISU(2)$
as its (transitive) isometry group and can be identified with the coset $ISU(2)/SU(2)$\,.
As a result, the space of complex functions on $\mathbb E^3$\,, denoted $C(\mathbb E^3)$\,,
is the space of $SU(2)$-invariant functions on $ISU(2)$\,.
Let us recall that  $C(\mathbb E^3)$ is a commutative algebra
endowed with the point-wise product.
The non-commutative algebra  we want to construct is a deformation of $C(\mathbb E^3)$\,.

This section is decomposed as follows.
First, we recall the classical construction of $C(\mathbb E^3)$
from the convolution algebra $C(\mathbb R^3)^*$ of distributions on $\mathbb{R}^3$
 before going to the deformed case.
Then, we show that deforming $C(\mathbb R^3)^*$ and using
the Hopf algebra duality principle we get the
convolution algebra $C(SU(2))^*$ of distributions on $SU(2)$\,.

\subsection{Construction in the classical case}

 Here we recall the construction in the undeformed case where
 $\mathcal{D}(SU(2))$ is replaced by $\mathbb C[ISU(2)]$\,.
We introduce the dual of the coset
$\mathbb C[ISU(2)]/\mathbb C[SU(2)]=\mathbb C[\mathbb R^3]$
which is defined as
$SU(2)$-invariant sesquilinear forms on $\mathbb C[ISU(2)]\,$ with compact
support:
\be
    \left\{F\in \mathbb C[ISU(2)]^*\;\left|\;
    \begin{array}{l}
    \forall\,(f\otimes u)\in \mathbb C[ISU(2)]\,,\,
    \forall\, v\in SU(2)\,,\\
     \la (f\otimes u)\cdot(1\otimes v)\,,\,F\ra\equiv\la
     (f\otimes uv)\,,\,F\ra=\la (f\otimes u)\,,\,F\ra
    \end{array}
    \right.
    \right\},
\ee with $\la c_1\,a\,,\,c_2\,F\ra=\bar{c_1}\,c_2\,\la a\,,\,F\ra$
for any $c_1,c_2\in\mathbb{C}$\,. This set can be identified with
the space of distributions on $\mathbb R^3$\ of compact support as
follows: \be
    C(\mathbb R^3)^*
    =\left\{\,\psi\;|\;\la f\,,\,\psi\ra
    =\la (f\otimes e)\,,\,F\ra\,,\quad \forall \,f\in C(\mathbb{R}^3)\,\right\}.
\ee
Furthermore, $C(\mathbb R^3)^*$ is equipped with an algebra structure
obtained from the Hopf duality principle.
This principle  allows to  define a Hopf algebra structure to the dual $\cal H^*$
of a given Hopf algebra $\cal H$\,. In particular, $\cal H^*$ inherits an algebra structure
from the co-algebra structure of $\cal H$\,.
In our particular case, the product $\psi_1\circ \psi_2$ of two distributions
$\psi_1$ and $\psi_2$ of $C(\mathbb R^3)^*$ is given by
\be
    \la f\,,\,\psi_1\circ \psi_2\ra
    \equiv \la \,\Delta_{\mathbb{C}[ISU(2)]}(f,e)\,,\,F_1\otimes F_2\,\ra
    = \la \Delta_{C(\mathbb{R}^3)}(f)\,,\,\psi_1\otimes \psi_2\ra\,,
\ee
where $\Delta_{C(\mathbb{R}^3)}$ is the usual co-product on $C(\mathbb R^3)$
defined in eq.(\ref{C(R^3) coproduct}).
In fact, this product is the convolution product on $C(\mathbb R^3)^*$
since it can be written as the standard definition of the convolution product
on the space of distributions:
\be\label{undef convol}
    \la f\,,\,\psi_1\circ \psi_2\ra\,= \la \la f\,,\,T[\psi_2]\ra\,,\,\psi_1 \ra\,,
\ee
where $T[\psi]$ is the $C(\mathbb R^3)$-valued distribution defined by
$\la f\,,\,T[\psi](p)\ra=\la T_{p} f \,,\,\psi \ra$
and $T_{p}$ is the translation operator: $(T_{p} f)(q)=f(p+q)$\,.

To complete the $\mathbb{C}[ISU(2)]$ case study, we introduce the
Fourier transform on $C(\mathbb R^3)^*$ and we get a  subspace of
functions on the Euclidean space $\mathbb E^3$\,, $C(\mathbb{E}^3)$\, whose Fourier transform are of
compact support.
 Indeed, the
Fourier map is defined as follows:
\ba
    \mathfrak{F}\,:\,C(\mathbb{R}^3)^*
    &\longrightarrow &
    C(\mathbb E^3)\,,\nn
    \psi &\longmapsto & \mathfrak{F}[\psi]=\la
    T_x\,,\,\psi\ra\,,
    \label{FFourier}
\ea with $T_x(p)=e^{-i\,\vec p\cdot\vec x}$\,. We see that the Fourier transform
of a distribution $\psi$ reduces to its evaluation on a pure
translational element. Furthermore, its co-product is simply
$\Delta_{C(\mathbb{R}^3)}(T_x)=T_x\otimes T_x$ and then the
product of two elements $\Psi_1$ and $\Psi_2$ is given by \be
    (\Psi_1\circ \Psi_2)(x)
    \equiv \la T_x\,,\,\mathfrak{F}^{-1}(\Psi_1)\circ \mathfrak{F}^{-1}(\Psi_2)\ra
    =\Psi_1(x)\, \Psi_2(x) \,.
\ee

An important consequence of this construction is the action of the
symmetry group. The action of an element $(T_y,u)$ of $ISU(2)$ on
$C(\mathbb E^3)$ is induced by its action on $\mathbb C[ISU(2)]^*$
as follows:
\ba\label{undef action}
    ((T_y\otimes u)\rhd \Psi)(x)&\!\!\equiv\!\!&
    \la(T_x\otimes e)\,,\, (T_y\otimes u)\rhd F\ra\equiv
    \la (T_y\otimes u)^*\cdot(T_x\otimes e)\,,\,F\ra
    \nn
    &\!\!=\!\!&
    \Psi(R(u^{-1})(\vec x-\vec y))\,.
\ea
Thus, one recovers the usual action of the Euclidean group.

\subsection{Construction in the deformed case}

Now, we go back to the deformed case of $\mathcal{D}(SU(2))$\,:
the construction follows the same steps.
We consider the dual of the coset $\mathcal{D}(SU(2))/\mathbb C[SU(2)]$
which replaces its undeformed counter part,
the dual of $\mathbb{C}[ISU(2)]/\mathbb C[SU(2)]$\,:
\be
    \left\{F\in \mathcal{D}(SU(2))^*\,\left|\,
    \begin{array}{l}
    \forall\,(f\otimes u)\in \mathcal{D}(SU(2))\,,\,\forall\,v\in SU(2)\,,\\
    \la (f\otimes u)\cdot(1\otimes v)\,,\,F \ra=\la (f\otimes u)\,,\,F \ra
    \end{array}
    \right. \right\}.
\ee
This space can be identified with the space of distributions on $SU(2)\,$:
\be
    C(SU(2))^*=\left\{\,\phi\;|\;
    \la f\,,\,\phi\ra=\la (f\otimes e)\,,\,F\ra\,,\quad \forall\,f\in C(SU(2))\,\right\},
\ee
and the algebra structure of $C(SU(2))^*$ is
obtained, as in the $\mathbb{C}[ISU(2)]$ case,
from the Hopf duality procedure.
The product of two elements $\phi_1$, $\phi_2$ of $C(SU(2))^{*}$ is given by
\be
    \la f\,,\,\phi_1\star \phi_2\ra
    \equiv \la \Delta_{\mathcal{D}(SU(2))}(f\otimes e)\,,\,F_1\otimes F_2\ra
    =\la \Delta_{C(SU(2))}(f)\,,\,\phi_1\otimes \phi_2\ra\,,
    \label{star_definition}
\ee
where $\Delta_{C(SU(2))}$ is the co-product defined in eq.(\ref{C(SU(2)) coproduct}).
One recognizes that this product is the usual convolution product on $C(SU(2))^{*}$
by recasting the above equation as follows:
\be
    \la f\,,\,\phi_1\star \phi_2\ra
    =\la\la f\,,\,L[\phi_2]\ra\,,\,\phi_1\ra\,,
    \label{def6}
\ee
where $L[\phi]$ is a $C(SU(2))$-valued distribution
defined by $\la f\,,\,L[\phi](u)\ra=\la L_uf\,,\,\phi\ra$
and the $L_u$ is the left deformed translation:
$(L_uf)(v)=f(uv)$\,.

The symmetry action of $\mathcal{D}(SU(2))$ on $C(SU(2))^*$
is induced by its action on $\mathcal{D}(SU(2))^*$\,.
The action of an element $(g\otimes u)$ of
$\mathcal{D}(SU(2))$ on $\phi \in C(SU(2))^*$ is defined by
\be\label{actionofDSU2}
    \la f\,,\,(g\otimes u)\rhd \phi\ra
    \equiv
    \la (f\otimes e)\,,\,(g\otimes u)\rhd F\ra
    \equiv
    \la (g\otimes u)^*\cdot (f\otimes e)\,,\, F\ra\nn
\ee
and in particular the action of
$C(SU(2))\subset\mathcal{D}(SU(2))$ is just the multiplication by
the  $C(SU(2))$ element: $\mathop{(g\otimes e)\rhd \phi=g\,\phi}$\,. The
$\mathcal{D}(SU(2))$ action on the product $\phi_1\star \phi_2$ is
defined using the co-algebra structure by
\be
    (g\otimes u)\rhd(\phi_1\star \phi_2)\equiv
    \sum_{(g)}\left((g_{(1)}\otimes u)\rhd\phi_1\right)\star \left((g_{(2)}\otimes u)\rhd\phi_2\right),
\ee
where we used the Sweedler notation for the co-product of ${C(SU(2))}$\,.

To finish, let us note that the algebra $C(SU(2))^*$ possesses an
antimorphic involution $\phi \mapsto \phi^{\flat}$ which plays the
role of the complex-conjugation. It is explicitly defined by
\be\label{flat}
	 \la f\,,\,\phi^{\flat} \ra \equiv \la S(f\otimes e)^*\,,\,F \ra\,, 
\ee 
using previous notations.

In order to get an intuition about the previous construction let us consider
some examples.

Of particular interest is the subalgebra of delta distributions.
The delta distributions $\delta_u \in C(SU(2))^*$
are defined for any $u\in SU(2)$ by
\be
    \la f\,,\,\delta_u\ra\,\equiv\,\overline{f(u)}\,,
\ee
and these delta distributions are in fact the eigenfunctions
of the momentum elements $(P_a \otimes e)$ of $\mathcal{D}(SU(2))$\,,
namely
$ (P_a\otimes e)\rhd\delta_u={P}_a(u)\,\delta_u$\,.
As a result,
$\delta_u$ can be interpreted as a pure momentum state of momentum $P_a(u)$\,.
The product (\ref{star_definition}) between two such
distributions  reads
\be
    \delta_{u_1}\star \delta_{u_2}=
    \delta_{u_1u_2}\,,
    \label{star delta}
\ee
giving the composition law of momenta recalled in the
Introduction.

We can now introduce the  coordinate distributions $\chi^a$. 
They are defined as the elements of the algebra on which the
infinitesimal translation, $((1-ix^b{P}_b)\otimes e)$\,, in the direction $x^b$\,
acts as \be
    ((1-ix^a{ P}_a)\otimes e)
    \rhd\,\chi^b =(1-ix^a{ P}_a)\,\chi^b=\chi^b-x^{b}\delta_e\,.
\ee
This equality can be recasted using a left-invariant vector
field $\xi_a$ (whose normalization is  implicitly given in the eq.(\ref{norm of chi}) below) 
as follows: \be\label{chidef}
    {P}_a(u)\left(\chi^b+2i\,{\ell_P}\,
    \xi^b\delta_e\right)=0\,.
\ee
The  solution for $\chi^a$ which is odd under  parity, $u\to u^{-1}$, is given by
\be
    \chi^a=-2i\,{\ell_P}\,\xi^a\delta_e\,.
\ee
When acting on a test function $f$, we have
\be\label{norm of chi}
    \la f\,,\,\chi^a\ra=-2i\,{\ell_P}\,\xi^a\,\overline f\,|_{e}=
        -2i\,{\ell_P}\frac{d}{d\theta}
    \overline{f\left(e^{-i\,\theta\,\mathcal J_a}\right)}\vert_{\theta=0}\,.
\ee
Notice that the sub-algebra generated by the $\chi^a$ is given by
the distributions whose support reduces to the identity element.
According to a theorem by L. Schwartz \cite{schwartz} the latter
sub-algebra is isomorphic to the universal enveloping algebra of
$\mathfrak{su}(2)$\,, ${\cal U}(\mathfrak{su}(2))$\,, the image of
$\chi^a$ being the generators of $\mathfrak{su}(2)$\,
and then satisfy
\be\label{chialgebra}
    [\chi^a ,\chi^b]_\star \equiv \chi^a\star \chi^b - \chi^b \star \chi^a
    = 2i\,\ell_P\, \epsilon_{abc} \, \chi^c \,.
\ee 
This relation is a consequence of  the relation giving the
product of two coordinates: \be
    \la f\,,\, \chi^a \star \chi^b \ra =
    -4\,{\ell_P}^2 \,\xi^a\,\xi^b\,\overline f \,|_{e}=
    -4\,{\ell_P}^2 \frac{\partial^2}{\partial \theta\partial \theta'}
    \overline{f\left(e^{-i\,\theta\,\mathcal J_a}\,e^{-i\,\theta'\,\mathcal J_b}\right)}
    \vert_{\theta=\theta'=0}
\ee
and the standard commutation relation of $\mathfrak{su}$(2).

Another interesting subalgebra of $C(SU(2))^*$ is the algebra of
functions $\phi$ on $SU(2)$ viewed as distributions in the
standard way: \be
    \la f\,,\,\phi\ra
    = \int d\mu(u)\,\overline{f(u)}\,\phi(u)\,,
\ee where $d\mu(u)$ is the Haar measure on $SU(2)$ with the
normalization $\int d\mu(u)=1$\,. The product between two such
elements is the convolution product on $C(SU(2))$\,: \be
    (\phi_1\star \phi_2)(u)=\int d\mu(v)\,\phi_1(v)\,\phi_2(uv^{-1})\,.
\ee

We end this paragraph by noting that when restricted to the space
of functions $C(SU(2))$\,, the non-commutative algebra we
constructed is equipped with a Hermitian scalar product defined
from the $SU(2)$ Haar measure. Given two functions $\phi_1$ and
$\phi_2$, their scalar product is given by the sesquilinear form
$\la\phi_1\,,\,\phi_2\ra$\, where $\phi_2$ is viewed as a
distribution. The action (\ref{actionofDSU2}) is unitary with
respect to this Hermitian form, namely, for any element $(g\otimes
u)$ in $\mathcal{D}(SU(2))$\,, it satisfies \be
    \la \phi_1\,,\,(g\otimes u)\rhd\phi_2 \ra=
    \overline{\la \phi_2\,,\,(g\otimes u)^*\rhd\phi_1\ra}\,.
\ee

\section{The non-commutative algebra $C_{\ell_P}(\mathbb{E}^3)$}

This section aims at linking, having physical applications in
mind, the sets $C(SU(2))^*$ and $C(\mathbb E^3)$ viewed, first, as
vector spaces. In other words, we want to interpret any
distribution $\phi \in C(SU(2))^*$ in terms of functions on the
classical Euclidean space $\mathbb E^3$\,. To do so, it is in fact
sufficient to make a link between $C(SU(2))^*$ and $C(\mathbb
R^3)^*$ and then to make use of the classical Fourier map
$\mathfrak F$ to go from $C(\mathbb R^3)^*$ to $C(\mathbb E^3)$\,.
Then, we extend these linear maps between $C(SU(2))^*$ and
$C(\mathbb R^3)^*$ or $C(\mathbb E^3)$ to morphisms in order to
endow the two later spaces with non-commutative products. The
obtained algebras will be respectively denoted by
$C_{\ell_P}(\mathbb R^3)^*$ and $C_{\ell_P}(\mathbb E^3)$: they
are deformations of $C(\mathbb R^3)^*$ and $C(\mathbb E^3)$ with
deformation parameter $\ell_P$.

\subsection{Relation between $C(SU(2))^*$ and $C(\mathbb R^3)^*$}

Let us concentrate on the link between $C(SU(2))^*$ and $C(\mathbb R^3)^*$\,.
As we have already emphasized in  Section 2.2.
there is no one-to-one mapping between $SU(2)$ and (subsets of) $\mathbb R^3$
because they are not homeomorphic.
As a result, it is impossible to find a one-to-one mapping
between $C(SU(2))^*$ and $C(\mathbb R^3)^*$\, and the construction
of a map between these two spaces is rather involved.

Our starting point is the mapping $\varphi$ which provides a mapping from $SU(2)$ to $\mathbb R^3$\,.
The map $\varphi$ is characterized by a function $\Pi$ that we will take to be one for simplicity.
As we have already emphasized there is no way to find a bijection 
between $C(\mathbb R^3)$ and any subset of $C(SU(2))$
for topological reasons; then
this prevents $\varphi$ from being an
injective map from the whole $SU(2)$ to $\mathbb R^3$.
However, it becomes injective when restricted to some subsets of $SU(2)$.
In that case, $\varphi$ allows to construct the following 3 bijections
from subsets of $SU(2)$ to subsets of $\mathbb{R}^3$\,:
\ba
    P_\pm \;\; : & U_\pm &\longrightarrow \quad B_{\ell_P} \,,\nn
    P_0 \;\; : & U_0\simeq S^2 &\longrightarrow \;{\partial B_{\ell_P}}\simeq S^2 \,,
\ea 
where $U_+$\,, $U_-$ and $U_0$ are respectively the northern
hemi-sphere, southern hemi-sphere and equator of $S^3\simeq SU(2)$
and $B_{\ell_P}$ is the open ball of $\mathbb R^3$, ${\partial
B_{\ell_P}}$ its boundary: 
\ba
    & U_\varepsilon  \!\!\!&= \left\{u(\vec{y},y_4)\in SU(2)
    \;\vert\;\text{sgn}(y_4)=\varepsilon\right\}\,,\nn
    & B_{\ell_P} \!\!\!& = \left\{ \vec p \in \mathbb{R}^3
    \;\vert\; |\,\vec p\,|<{\ell_P}^{-1} \right\}\,,
\ea
with $\text{sgn}$ being the sign function such that \m{\text{sgn}(0)=0}\,.
As a consequence, we  have a natural decomposition of $C(SU(2))^*$ into
$C_{U_+}(SU(2))^*$\,, $C_{U_-}(SU(2))^*$ and $C_{U_0}(SU(2))^*$ where
$C_V(SU(2))^*$ denotes the space of $SU(2)$-distributions with support on $V\subset SU(2)$\,;
since $U_\pm$ are open subsets, $C_{U_\pm}(SU(2))^*$ are in fact
the spaces $C(U_\pm)^*$ of distributions on $U_\pm$\,. As a result, we have the following identification:
\be
C(SU(2))^* \simeq C(U_+)^* \oplus C(U_-)^* \oplus C_{U_0}(SU(2))\* \,,
    \qquad \phi \simeq
    \phi_+\oplus\phi_-\oplus\phi_0\,,\label{phidecomp}
\ee 
where the components $\phi_\pm$ are explicitly obtained from
the characteristic functions $I_\pm$ on $U_\pm$ by the formulae
$\phi_\pm=I_\pm\,\phi$ \footnote{ To explicitly define the
characteristic functions $I_\pm$ on $U_\pm$\,, let us consider the
limit of a test function $I^\epsilon \in C^\infty(\,[-1,1]\,)$
with $\epsilon <1$ defined by the fact that, $I^\epsilon(y_4)=1$
for $y_4>\epsilon$ and $d^s/dy_4^s I^\epsilon(0)=0$ for $s\ge
0$\,. As a result, $\phi_\pm$ are given by the relations: \be
    I_\pm\,\phi = \lim_{\epsilon \to 0} I^\epsilon(\pm y_4)\,\phi\,.
\ee }. The remaining component $\phi_0$ is obtained by difference:
$\phi_0=\phi-\phi_+-\phi_-$. We can characterize more explicitly
the space $C_{U_0}(SU(2))^*$. Contrary to the other cases, this
space does not reduce to the space $C(U_0)^*$ of distributions on
$U_0$ but decomposes into an infinite sum of them. More precisely,
any $\phi_0\in C_{U_0}(SU(2))^*$ decomposes as follows: \be \phi_0
\; = \; \sum_s \phi_{0s} \, \delta^{(s)}(y_4) \ee where the sum is
finite, $\delta^{(s)}(y_4)=d^s\delta(y_4)/(dy_4)^s$ and $\phi_{0s}
\in C(U_0)^*$ which is obtained from $\phi_0$ by the integral: 
\be
    \phi_{0s}(\vec n) = \frac{1}{s!}\int_{-1}^1 dy_4\,(-y_4)^s\,\phi_0(\vec n,y_4)\,.
\ee
For later convenience, we introduce the \emph{polarization} vectors $\xi_\varepsilon$ with
$\varepsilon\in\{+,-,0\}$ defined by $\phi \simeq \phi_\varepsilon\, \xi_\varepsilon\,$, i.e.
\be
    \xi_+=1\oplus0\oplus 0, \quad
    \xi_-=0\oplus1\oplus 0, \quad
    \xi_0=0\oplus0\oplus 1 \,.
\ee
It is also convenient to introduce the dual polarization vector $\xi_\varepsilon^t$ defined by
the relation $\xi_\varepsilon^t\, \xi_{\varepsilon'}=\delta_{\varepsilon\varepsilon'}$.

The idea is now to map the spaces $C(U_\pm)^*$ and $C_{U_0}(SU(2))^*$
to  spaces of distributions on
$\mathbb R^3$. This will make the link between the curved
space of momenta and the ordinary flat momentum space in 3 dimensions. More precisely, we look for
the following linear mappings:
\ba
    \mathfrak{a}_\pm \;\; : & C(U_\pm)^* &\longrightarrow
    \;\; C_{B_{\ell_P}}(\mathbb{R}^3)^*\,,\nn
    \mathfrak{a}_{0} \;\; : & C_{U_0}(SU(2))^* &\longrightarrow
    \;\; C_{\partial B_{\ell_P}}(\mathbb{R}^3)^*\,.
\ea
For purposes of simplicity, let us define a mapping $\mathfrak{a}$ as a multiplet of the above mappings:
$\mathfrak{a}\equiv\mathfrak{a}_+\oplus\mathfrak{a}_-\oplus \mathfrak{a}_{0}$\,,
and denote the image of $C(SU(2))^*$ by $\mathfrak{a}$ as $C_{\ell_P}(\mathbb{R}^3)^*$\,:
\be
    \mathfrak{a}\;:\; C(SU(2))^* \;\longrightarrow\;
    C_{\ell_P}(\mathbb{R}^3)^*\equiv
    C_{B_{\ell_P}}(\mathbb{R}^3)^*\oplus C_{B_{\ell_P}}(\mathbb{R}^3)^*
    \oplus  C_{\partial B_{\ell_P}}(\mathbb{R}^3)^*\,.
\ee
Note that $C_{\ell_P}(\mathbb{R}^3)^*$ is interpreted as a deformation of the space $C(\mathbb{R}^3)^*$
of distributions on $\mathbb R^3$.
We require, in addition, that
the action of the Poincar\'e group $ISU(2)\subset \mathcal{D}(SU(2))$
induced by the mapping $\mathfrak{a}$
on each component of $C_{{\ell_P}}(\mathbb{R}^3)^*$
is the standard covariant one (\ref{undef action})\,.
Therefore, we have the following conditions:
\be
    (T_x\otimes u)\rhd\mathfrak{a}(\phi)
    \equiv \mathfrak{a}(\varphi(T_x\otimes u)\rhd\phi)
    \equiv \mathfrak{a}(\varphi(T_x)\,\text{ad}_{u}(\phi))
    = T_{x}\,\text{R}_{u}(\mathfrak{a}(\phi))\,.
\ee
Note that each components $C(U_\pm)^*$ and $C_{U_0}(SU(2))^*$ of $C(SU(2))^*$
are stable under the action of $\mathcal D(SU(2))$\,.
As a result, the solutions of the above condition for $\mathfrak{a}_\pm$ are
\be
    \mathfrak{a}_\pm(\phi_\pm) =  g_\pm\,\varphi_*(\phi_\pm)\,,
\ee
where $g_\pm$  are functions of the norm $|\,\vec p\,|$ and
$\varphi_*$ is the pull-back of $\varphi$ on the space of distributions
$C(SU(2))^*$ defined by
\be
    \forall f\in C(\mathbb{R}^3)\,,\quad
    \la f\,,\,\varphi_*(\phi)\ra=\la \varphi(f)\,,\,\phi\ra\,.
\ee
For simplicity, we make the choice $g_\pm=1$.

We can proceed in the same way to find the general solution for
$\mathfrak a_0$. Unfortunately, with the same choice of $\varphi$,
related to $\vec P(u)$, $\mathfrak a_0$ admits a non-trivial
kernel and then is not a bijection: for example, $\mathfrak
a_0(\delta'(\theta -\pi))=0$ where we used the parametrization
$(\theta,\vec n)$ for $SU(2)$. This is  due to the fact that the
derivative of $|\vec P|(\theta)$  vanishes for $\theta=\pi$, i.e. on
$U_0$. To construct a bijective mapping $\mathfrak a_0$, it is
necessary to make a new choice $\vec P_0(u)=P_0(\theta)\,\vec n$
around $U_0$ which induces a new mapping $\varphi_0$\,. More
precisely, we consider an open set of $SU(2)$ containing $U_0$ and
we require $P_0(\theta)$ strictly monotonous with
$P_0(\pi)=P(\pi)={\ell_P}^{-1}$. Then, the solution of the above
condition for $\mathfrak a_0$ leads to 
\be
    \mathfrak{a}_0(\phi_0) =  g_0\,\varphi_{0*}(\phi_0)\,,
\ee
where $g_0$ is a constant; we will make the choice $g_0=1$ for simplicity.
Note that the choice of $\vec P_0$ is arbitrary and therefore there exist
ambiguities in the definition of the mapping $\mathfrak a_0$\,. 
In fact, these ambiguities are similar to those
we have already discussed concerning the choice of the momentum coordinates in the deformed theory.

Using the usual Fourier transform $\mathfrak{F}$ defined in
eq.(\ref{FFourier}), one maps any multiplet of distributions in
$C_{\ell_P}(\mathbb{R}^3)^*$ into a multiplet of functions on
$\mathbb E^3$. The image of $C_{\ell_P}(\mathbb{R}^3)^*$ is
denoted $C_{\ell_P}(\mathbb{E}^3)$: it is interpreted as a
deformation of the space $C(\mathbb E^3)$ of classical functions
on $\mathbb E^3$. It will be convenient to denote by $\mathfrak
m\equiv \mathfrak{F}\circ \mathfrak{a}$ the mapping from
$C(SU(2))^*$ to $C_{\ell_P}(\mathbb{E}^3)$\, and $\mathfrak m$
defines a bijection between $C(SU(2))^*$ and the space
$C_{\ell_P}(\mathbb{E}^3)$. The image of an element $\phi$ in
$C(SU(2))^*$ by the mapping $\mathfrak{m}$ is explicitly given by
\be\label{relation Phiphi}
    \mathfrak{m}(\phi)(x) = \la e^{-i\,\vec P\cdot \vec x}\,,\,\phi_+ \ra \oplus
    \la e^{-i\,\vec P\cdot \vec x}\,,\, \phi_-\ra \oplus
    \la e^{-i\,\vec P_0\cdot \vec x}\,,\,\phi_0\ra
    \equiv \Phi(x)
\ee

As a vector space, \m{C_{\ell_P}(\mathbb E^3)\simeq \widetilde{ C}_{\ell_P}(\mathbb{R}^3)^*
=\widetilde{C}_{B_{\ell_P}}(\mathbb{R}^3)^*
\oplus \widetilde{C}_{B_{\ell_P}}(\mathbb{R}^3)^*
\oplus  \widetilde{C}_{\partial B_{\ell_P}}(\mathbb{R}^3)^*}
where $\widetilde C$ is the Fourier image of the vector space $C$\,.
The space  $C_{\ell_P}(\mathbb E^3)$ inherits a non-commutative algebra structure we will describe in the next Section.

To illustrate the previous notions, let us consider some clarifying examples.

\subsubsection*{Example 1: functions on $SU(2)$}

First, let us assume that $\phi$ is a function, i.e. we restrict $C(SU(2))^*$  to the sub-algebra of
functions $C(SU(2))$: in that case,  $\phi_\pm$ are both functions
and $\phi_{0}=0$ in the decomposition of $\phi$\,.
The images of $\phi_\pm\in C(SU(2))$ by $\mathfrak{a}_\pm$ are
 completely determined by $\vec{P}(u)$ and given by
\ba\label{mapping a}
        \mathfrak{a}_\pm(\phi_\pm)(\vec{p}) \e
    \int d\mu(u) \,\delta^3(\vec p-\vec P(u))\,\phi_\pm(u)\nn
    \e \frac{v_{{\ell_P}}}{\sqrt{1-(\ell_P |\,\vec p\,|)^2}}\,\,
    \phi_\pm\Big(u\big(\ell_P\,\vec p,\pm\sqrt{1-(\ell_P |\,\vec p\,|)^2}\big)\Big)\,,
\ea
where we have introduced the constant of volume dimension $\mathop{v_{{\ell_P}}={\ell_P}}^3/(2\,\pi^2)$\,.

Furthermore, the Fourier transform of $\mathfrak a_\pm(\phi_\pm)$ is a function of $x$ which is related to
the function $\phi \in C(SU(2))$ by the integral:
\be\label{relPhiphi}
	\Phi_{\pm}(x)  = {\mathfrak m}_\pm(\phi_\pm)(x) \, = \,
	\int d\mu(u) \, \phi_\pm(u) \, e^{i\,\vec P(u)\cdot\vec x} \,.
\ee
This transform is invertible and the inverse relation can be obtained performing the following classical
Lebesgue integral on $\mathbb R^3$:
\be
	\phi_\pm(u) \, = \, \sqrt{1 -(\ell_P P(u))^2} \, \int \frac{d^3x}{(2\pi)^3 \, v_{\ell_P}} \, \Phi(x) \,
	e^{i\,\vec P(u)\cdot\vec x}\, .
\ee
As we will see in the sequel, it is possible to inverse the relation (\ref{relPhiphi}) making use
of a non-commutative $\star$-product defined in Section 4.2.

\subsubsection*{Example 2: elements of $C_{U_0}(SU(2))^*$}

A second example is the case where $\phi$ has a support on $y_4=0$
that is  \m{\phi=\sum_s\delta^{(s)}(y_4)\,\phi_{s}} where $\phi_{s}$ are
functions on $U_0\simeq S^2$\,.
A short calculation shows that
the images of $\phi$ by the mapping $\mathfrak{a}_0$ is a distribution on $\mathbb R^3$ 
whose support is the two-sphere $\partial B_{\ell_P}$\,:
\be
    \mathfrak{a}_0(\delta^{(s)}(y_4)\,\phi_s)(\vec{p})
    = \frac{\phi_s(\vec p/p)}{4\pi\,p^2}\,
    \sin^2\!\left(\frac{\theta_0(p)}2\right) \frac{\theta_0'(p)}{\theta_0'({\ell_P}^{-1})}
    \left(\frac{1}{\theta_0'(p)}\frac{d}{dp}\right)^s\delta(p-{\ell_P}^{-1})\,,
\ee
where the function $\theta_0(p)$ is the inverse function of $P_0(\theta)$\,:
$P_0(\theta_0(p))=p$\,.

In the particular case where $P_0$ is linear,
let us say $P_0(\theta)={\ell_0}^{-1}(\theta-\pi)+{\ell_P}^{-1}$ for instance,
then the mapping $\mathfrak a_0$ reduces to:
\ba\label{deltastodeltas}
    &&\mathfrak{a}_0(\delta^{(s)}(y_4)\,\phi_s)(p) =
    \frac{\phi_s(\vec p/p)}{4\pi\,{\ell_0}^{s}\,p^2}\,
    \cos^2\!\left(\frac{\ell_0}{2}\,(p-\ell_P)\right)
    \delta^{(s)}\!\left(p-{\ell_P}^{-1}\right)\\
    &&\quad =\,\frac{\phi_s(\vec p/p)}{8\pi\,{\ell_0}^{s}\,p^2}
    \left[\delta^{(s)}\!\left(p-{\ell_P}^{-1}\right)
    +\sum_{n=0}^{\lfloor s/2\rfloor}
    \binom{s}{2n}\,(-1)^n\,{\ell_0}^{2n}\,\delta^{(s-2n)}\!\left(p-{\ell_P}^{-1}\right)\right].
    \nonumber
\ea
Thus, the image of $\delta^{(s)}(y_4)$ is a sum of several $\delta^{(k)}(p-\ell_P{}^{-1})$ with $k\leq s$.

\subsubsection*{Example 3: delta distributions}

Another simple but important example is when $\phi$ is a delta distribution $\delta_{u}$\,, 
i.e. a plane wave.
A trivial calculation yields for any $u\in U_\varepsilon$ with $\varepsilon=+,-,0$
\ba
    \mathfrak a (\delta_u)(\vec{p}) = \delta^3(\vec p-\vec P(u))\,\xi_\varepsilon\,.
\ea
Note that $\vec P_0(u)=\vec P(u)$ for $u \in U_0$; for that reason, we write the same formula for
each $\varepsilon$.

It is now possible to compute the Fourier transform of these
distributions to obtain the analoguous of the classical expressions
of the plane waves. A simple calculation leads to the following
expressions:
\be\label{plane wave of x}
    w_u(\vec{x}) \equiv \mathfrak m(\delta_u)(\vec x)
    = e^{i\,\vec{P}(u)\cdot \vec{x}}\,\xi_\varepsilon\,.
\ee
Formally, the plane waves have the same expression as the classical ones.

\subsubsection*{Example 4: coordinate distributions}

Of particular interest are the coordinate distributions. Their
expressions in $C_{\ell_P}(\mathbb R^3)^*$ or $C_{\ell_P}(\mathbb
E^3)$ are also easy to obtain. Indeed, we have shown that, in the
$C(SU(2))^*$ representation, coordinates are the left-invariant
vector fields $\chi^a$ (\ref{chidef}) which admit only one
non-trivial component in the decomposition (\ref{phidecomp}),
$\chi^a_+=\chi^a$\,, because they are distributions localized at
the origin $e$. The images by $\mathfrak a$ and $\mathfrak m$ are
therefore singlet as well respectively given by
\be
    \mathfrak a(\chi^a)=i\frac{d}{dp_a}\delta^{3}(\vec{p})\,\xi_+\,,\qquad
    \mathfrak m(\chi^a)=x_a\,\xi_+\,.
\ee
These expressions agree with their classical counterpart.

\subsection{The $\star$-product}

Now the idea is to require, in some sense we will precise in the sequel,
the mapping $\mathfrak{a}$ to be algebra morphisms.
The product between two elements
$\psi_1$ and $\psi_2$ in $C_{\ell_P}(\mathbb{R}^3)^*$ is induced
from that of $C(SU(2))^*$ as follows
\be\label{psi star}
    \psi_1\star \psi_2=\mathfrak{a}\left({\mathfrak{a}}^{-1}(\psi_1)\star{\mathfrak{a}}^{-1}(\psi_2)\right).
\ee
This $\star$-product is far from being equal to the classical convolution product
$\psi_1\circ \psi_2$ presented in eq.(\ref{undef convol}) for two reasons.
First, $\psi_i$ are now triplets
$\psi_{i\,+}\oplus\psi_{i\,-}\oplus\psi_{i\,0}$
contrary to classical functions.
Second, even if $\psi_1$, $\psi_2$ and $\psi_1 \star \psi_2$
admit only one non-trivial component for each, let us say $\psi_{1+}$, $\psi_{2+}$ and
$(\psi_{1}\star \psi_{2})_+$, the resulting $\star$-product is no-longer the classical convolution product.

To be more precise, let us compute explicitly the product of two
plane waves, viewed as elements of $C_{\ell_P}(\mathbb R^3)^*$.
 A  plane wave is the image by $\mathfrak a$
of $\delta_u$ and  is thus characterized by the vector $\vec P(u)$
and the space $U_\varepsilon$ to which $u$ belongs. As a result, a
plane wave is given by $\delta^3_{\vec{p}}\,\xi_\pm$ with $|\,\vec
p\,|<{\ell_P}^{-1}$ or by $\delta^3_{\vec{p}}\,\xi_0$ with
$|\,\vec p\,|={\ell_P}^{-1}$\,. A short calculation leads to
the following product of two plane waves: \be\label{prodofdelta}
    \delta^3_{\vec{p}}\;\xi_\epsilon \;\star\;
    \delta^3_{\vec{q}}\;\xi_{\zeta} =
    \delta^3_{\vec{k}}\;\xi_{\eta}\,,
\ee
where
\ba\label{sumrule1}
    \vec k \eq
    \zeta\sqrt{1-(\ell_P\,|\,\vec q\,|)^2}\,\vec p +
    \epsilon\sqrt{1-(\ell_P\,|\,\vec p\,|)^2}\,\vec q +
   {\ell_P}\,\vec p \wedge \vec q\,,\nn
    \eta \eq \text{sgn}\left(\epsilon\zeta
    \sqrt{1-(\ell_P\,|\,\vec p\,|)^2}\sqrt{1-(\ell_P\,|\,\vec q\,|)^2}
    -{\ell_P}\, \vec p \cdot \vec q\right) \,.
\ea
In the dual point of view, the plane waves represent deformed momenta and their products
define in fact an addition rule of momenta.

The first important difference with the classical theory is that a
\emph{deformed} momentum is now characterized by a couple $(\vec
p,\varepsilon)$ where $\vec{p}$ is bounded by ${\ell_P}^{-1}$
and $\varepsilon$ is a discrete internal variable. Second, the
addition rule is deformed compared to its classical counterpart
and the resulting momentum vector depends on the initial
polarization vectors. Nevertheless, this addition rule agrees with
the classical one at the classical limit where $\ell_P$ goes to
zero and the parameter $\varepsilon$ is fixed to the value 1.

Now, let us see how this $\star$-product is expressed in the position representation.
The Fourier map $\mathfrak F$ induces an algebraic structure on $C_{\ell_P}(\mathbb E^3)$ and
we keep the notation $\star$ for the product between two elements of $C_{\ell_P}(\mathbb E^3)$\,.
The $\star$ product between two plane waves, viewed as elements of $C_{\ell_P}(\mathbb E^3)$ reads
\be
    e^{i\,\vec{p}\cdot \vec{x}}\,\xi_\epsilon\;\star\;
    e^{i\,\vec{q}\cdot \vec{x}}\,\xi_\zeta =
    e^{i\,\vec{k}\cdot \vec{x}}\,\xi_\eta\,,
\ee
where $\vec k$ and $\eta$ are given by the formulae (\ref{sumrule1}).

Now, let us compute the $\star$ product of two elements
\m{\psi_1,\psi_2 \in C_{\ell_P}(\mathbb R^3)} which are the images by $\mathfrak a$ of two functions
on $SU(2)$, $\phi_1$ and $\phi_2$.
In other words, $C(SU(2))^*$ is restricted to the sub-algebra of functions $C(SU(2))$\,,
and we study the algebraic properties of the image of $C(SU(2))$ by $\mathfrak a$.
First, we remark that $C(SU(2))$ is stable by the convolution product.
Then, we recall that the only non-vanishing components of $\phi \in C(SU(2))$ are $\phi_+$ and $\phi_-$.
Therefore, the image of $\phi$ consists only in a couple of functions
given, in terms of delta distributions, by
\be\label{devdelta}
    \psi = \sum_{\epsilon=\pm}
    \int_{B_{\ell_P}} d^3\vec p\;\psi_\epsilon(\vec p)\,\delta^3_{\vec p}\;\xi_\epsilon\,.
\ee

To compute the $\star$ product $\psi_{1}\star \psi_{2}$\,,
one uses the above decomposition 
in terms of delta distributions and making use of the formula (\ref{prodofdelta}) of
the $\star$-product between delta distributions, one obtains
\ba
    \psi_{1}\star \psi_{2} = \sum_{\epsilon,\,\zeta=\pm}
    \int_{B_{\ell_P}^{\;\times 2}} d^3\vec p \, d^3\vec q \;
    \psi_{1\,\epsilon}(\vec p) \,\psi_{2\,\zeta}(\vec q)\;
    \delta^3_{\vec k}\;\xi_\eta\,,
\ea
where $\vec k$ and $\eta$ are given by (\ref{sumrule1}).
At the classical limit, this $\star$-product
reduces to the classical convolution product on $\mathbb R^3$\,.

The application $\mathfrak F$ maps $\psi_+\oplus\psi_-\oplus0\in C_{\ell_P}(\mathbb R^3)$ to a pair
of functions \m{\Phi_+\oplus\Phi_-\oplus0\in C_{\ell_P}(\mathbb E^3)}
 where $\Phi_\pm$ are functions of the variable $x$\,:
\ba
    \Phi_\pm(x) \e  \int_{U_\pm} d\mu(u) \, \phi(u) \, e^{i\,\vec P(u)\cdot \vec x} \nn
    \e \int \frac{d^3\vec y}{2\sqrt{1-|\vec y|^2}} \,
    \phi(\vec y,\pm\sqrt{1-y^2})\, e^{i\,{\ell_P}^{-1}\vec x \cdot \vec y}\,,
\ea 
where we have identified $u$ with $(\vec y,y_4)$ where the
integral is defined over the vector $\vec y$ such that $y<1$.
Thus, the image of $C(SU(2))\subset C(SU(2))^*$ by $\mathfrak
m=\mathfrak F \circ \mathfrak a$ is the sub-space of
$C_{\ell_P}(\mathbb E^3)$ where the third component is null.
The product between two such functions $\Phi_1$ and $\Phi_2$ is
implicitly given by \be
    \Phi_1  \star \Phi_2 =
    \mathfrak m \left(\mathfrak m^{-1}(\Phi_1) \circ \mathfrak m^{-1}(\Phi_2)\right) \;.
\ee
In fact, the product is a couple of functions.
The explicit expression of each term of the couple is neither needed nor simple but can be given
in terms of a non-local integral of the form:
\be
    (\Phi_1  \star \Phi_2)(x)
    = \sum_{\epsilon,\zeta=\pm}\int d^3y \, d^3z \,
    \mathcal K_{\epsilon\,\zeta}(x;y,z) \,
    \Phi_{1\,\epsilon}(x_1)\,\Phi_{2\,\zeta}(x_2)\,,
\ee
where the kernel $\mathcal{K}_{\epsilon\,\zeta}$ is defined by the double integral
\be
    \mathcal{K}_{\epsilon\,\zeta}(x;y,z)  =
    \int d^3\vec p \, d^3\vec q \,
    e^{i\,\left(\vec p \cdot \vec y + \vec q \cdot \vec z +\vec k \cdot \vec x\right)} \,\xi_{\eta}\,,
\ee
with $\vec k$ and $\eta$ are given by the sum rule
(\ref{sumrule1}). This formula simplifies drastically when
$\Phi_i$ are chosen to be coordinates functions. In that case, the
integral can be performed explicitly and one obtains
\be\label{coord algebra}
	x^a\,\xi_+ \star x^b\,\xi_+ = 
	\left (x^a\,x^b + {i}\,\ell_P\,\epsilon_{abc}\,x^c \right) \xi_+\,,
\ee 
which illustrates the non-commutativity of the coordinates. As expected
from (\ref{chialgebra}), one sees immediately that
\be\label{coord commuator}
	[x^a\,\xi_+,x^b\,\xi_+]_\star\equiv x^a\,\xi_+\star x^b\,\xi_+ -
	x^b \,\xi_+\star x^a \,\xi_+= 2i\,\epsilon_{abc}\,\ell_P\,x^c\,\xi_+\,,
\ee
which is the $\mathfrak{su}$(2) algebra.

\subsection{Invariant measure on $C_{\ell_P}(\mathbb{E}^3)$ and scalar action}

In order to construct a local action, it is necessary to define an invariant measure on $C(SU(2))^*$\,.
By invariant, we mean invariant under translations and rotations.
As in the undeformed case, there is no hope to define an invariant
measure on the whole distributional algebra $C(SU(2))^*$\,: we
will focus on the sub-algebra of functions $C(SU(2))$\,. 

From the very construction, elements of $C(SU(2))$
are linear forms on a subset of $\mathcal{D}(SU(2))$ which is,
itself, endowed with (a two dimensional space of) invariant measures in the sense of Hopf algebras.
Thus $C(SU(2))$ inherits naturally invariant measures which depends on $\alpha,\beta \in \mathbb C$ as
follows:
\ba
    h \,: \,& C(SU(2)) &\longrightarrow \quad \mathbb C \,,\nn
    &\phi \, &\longmapsto \quad h(\phi)=\alpha\,\phi(e)\,+ \, \beta\,\phi(e_A) \,.
\ea
It is immediate to check that $h$ is invariant under rotations and translations.
It is also easy to see that $h$ does not extend to the whole distributional algebra
$C(SU(2))^*$\,,
the delta distribution at $e$ or $e_A$ for instance being non-normalizable with respect to this measure.
Furthermore, $h$ allows to define a bilinear hermitian form on $C(SU(2))$ such that the \emph{scalar} product between
$\phi_1$ and $\phi_2$ is given by:
\be
	h(\phi_1^{\flat} \, \circ \, \phi_2)  =  \alpha \int d\mu(u) \, \overline{\phi_1(u)} \, \phi_2(u) \,
	+ \, \beta \int d\mu(u) \, \overline{\phi_1(u)} \, \phi_2(u\,e_A)
\ee
where we have used the notation $\phi^{\flat}=\overline{\iota(\phi)}$ introduced in eq.(\ref{flat}).
In order to have a positive definite
bilinear form, one has to assume that $\beta$ vanishes. We will make this assumption in the sequel, together with the
choice $\alpha=1$.

Having in mind the construction of a non-commutative QFT,
it is useful to export $h$ in the $C_{\ell_P}(\mathbb E^3)$
formulation of $C(SU(2))^*$ via the map $\mathfrak m$\,:
\be\label{measureClp}
    H(\Phi) \equiv h(\mathfrak{m}^{-1}(\Phi))=\int \frac{d^3x}{(2\pi)^3\,v_{\ell_P}} \, \xi^t_+\,\Phi(x)\,,
\ee
where the integral in the r.h.s. is defined from the standard Lebesgue measure
on $\mathbb R^3$. Let us remind that, as $\Phi=\mathfrak m (\phi)$ is the image of a function $\phi$,
it admits only two components $\Phi_\pm$ and then the measure $h$ (\ref{measureClp}) involves only one of these two
components. Let us note however that the norm of $\Phi$ involves its two components, namely:
\be\label{normPhi}
    h(\phi^{\flat} \circ \phi)  =  \xi_+^t
    \int \frac{d^3x}{(2\pi)^3 v_{\ell_P}}\big( \overline{\Phi_+(x)}\xi_+ \star \Phi_+(x)\xi_+  +
    \overline{\Phi_-(x)}\xi_- \star \Phi_-(x)\xi_- \big) \,.
\ee

Such a measure is necessary to define an action for a QFT in the non-commutative space
$C_{\ell_P}(\mathbb E^3)$. For instance, the free action for a scalar field
$\Phi=\mathfrak m(\phi)$, where $\phi$ is a function on $SU(2)$, is given by
the following integral:
\be
    \frac{2S[\phi]}{(2\pi)^3 \, v_{\ell_P}}=-h\left(\phi^\flat\circ(P^2+m^2)\phi\right)=
     h\left( (P\phi)^{\flat} \circ (P\phi)\right) - m^2 h(\phi^{\flat} \circ \phi)
\ee
where $P$ is the momentum function (\ref{def de P}). This action is constructed by analogy
with its classical \emph{undeformed} counterpart: it is quadratic, of second order and local according to the
$\star$-product in the position representation.
It is immediate to notice that the components $\Phi_+$ and $\Phi_-$
decouple and then, using (\ref{normPhi}), on shows that the action reads $S[\Phi]=S_+[\Phi_+] + S_-[\Phi_-]$ where:
\be
    S_\varepsilon[\Phi_\varepsilon]
    =\frac{1}{2}\xi_+^t \int {d^3x}
    \left( \partial^a \Phi_\varepsilon \xi_\varepsilon\star \partial_a
    \Phi_\varepsilon \xi_\varepsilon-m^2\,
    \Phi_\varepsilon \xi_\varepsilon\star \Phi_\varepsilon \xi_\varepsilon \right)\;.
\ee
We have assumed for simplicity that the fields $\Phi_\pm$ are real, i.e. $\overline{\Phi_\pm}=\Phi_\pm$ which is equivalent to the condition $\phi=\phi^{\flat}$\,.

Contrary to the Moyal case, the integral of the $\star$-product of
two functions is different from the classical integral.
To illustrate some
differences between the deformed and the classical integrals of
the product of two functions, let us consider the example where
the functions are plane waves. In the case of the $\star$-product,
one gets
\be
    \int \frac{d^3x}{(2\pi)^3\,v_{\ell_P}} \,(w_u\star w_v)(x)
    =\int \frac{d^3x}{(2\pi)^3\,v_{\ell_P}} \,w_{uv}(x)
    =\delta_{e}(uv)\,\xi_+\,,
\ee
whereas for the point-wise product, one gets
\be
    \int \frac{d^3x}{(2\pi)^3\,v_{\ell_P}} \,e^{i\,\vec P(u)\cdot \vec x}\,e^{i\,\vec P(v)\cdot \vec x}
    =\frac{1}{\sqrt{1 - (\ell_P\,P(u))^2}} \, \delta_{e}(uv)\,.
\ee

The difference between the non-commutative and standard commutative QFT is even deeper
when one considers self-interactions. As an example, let us consider a cubic self-interaction
which is defined such that the momenta addition rule is satisfied at each vertex. In the classical case,
this requirement leads to the standard action whereas, in the non-commutative case, it leads to
the following simple interacting term with coupling constant $\lambda$\,:
\be
    S^{(3)}[\phi] =(2\pi)^3\,v_{\ell_P} \,  \frac{\lambda}{3!} \, h(\phi \circ \phi \circ \phi),
\ee
in the momentum representation. The generalization to any polynomial interaction is immediate.
It is interesting to write this integral in terms of the functions $\Phi_{\pm}$ and, after some
calculations, one gets
\be
    S^{(3)}[\Phi] = \frac{\lambda}{3!}\,\xi_+^t
    \int {d^3x}
    \left(\Phi\star \Phi\star \Phi \right)(x)
\ee
In the momentum representation where 
$\psi={\mathfrak F}^{-1}[\Phi]$ this vertex reads
\be
	 S^{(3)}[\psi] = \frac{\lambda}{3!}
	\sum_{\epsilon,\zeta=\pm}\int d^3pd^3q\,
	\psi_\epsilon(p)\,\psi_\zeta(q)\,\psi_\eta(-k)\,,
\ee
where $\vec k$ and $\eta$ are related to $\vec p$, $\vec q$ and $\epsilon$, $\zeta$ by eq.(\ref{sumrule1}).
Contrary to the free action,
this interaction term couples the two components of the field $\Phi$\,:
there are four different vertices in the theory.
Only the case where
one out of the two components is non-trivial has been investigated so far.

We leave the precise study of this action for future
investigations but we can give some preliminary interesting
results. First, note that at each vertex the momentum conservation
holds with the deformed addition rule. Next, let us consider the
Feynman propagator of the free field theory which has been studied
in \cite{FL2}. In the momentum representation, it is simply given by 
\be
    \Gamma(u) = \frac{1}{P^2(u) + m^2}\,,
\ee
with $P(u) = {\ell_P}^{-1}\,\sin(\theta/2)$\,.
An immediate analysis leads to the fact
that the associated functions $G(x;x')$ depend
only on the distance $\vert \vec x-\vec x' \vert$ between two positions $x$ and $x'$\,,
and are given by the following equivalent expressions:
\be
    G(x;x')=\frac{1}{(2\pi)^3\,v_{\ell_P}} \, \int_{U_+} d\mu(u) \, \frac{e^{i\,\vec P(u)\cdot (\vec x - \vec x')}}{P^2(u) + m^2}
     =\frac{1}{8\,r}\,\int_0^1\frac{t\,dt}{\sqrt{1-t^2}}\,
    \frac{\sin({\ell_P}^{-1}\,r\,t)}{t^2+({m\,\ell_P})^2}\,,
    \label{GX}
\ee
with \m{r=|\vec x-\vec x'|}\,.
It is interesting to compare this function with its classical counterpart:
\be
    G_{cl}(x;x')  = \int \frac{d^3p}{(2\pi)^3}\,
    \frac{e^{i\,\vec p\cdot (\vec x-\vec x')}}{p^2+m^2} =
    \frac{\exp(-\,m\,r)}{4\pi\,r}\,.
\ee 
Contrary to the classical case, the integral over the momentum $p$
is definite and the upper bound depends on the Planck length
$\ell_P$. This important fact makes the Feynman propagator
well-defined at the coincident point limit $r\to 0$\,: \be
    G(x;x)=G(0;0)=\frac{1}{4\pi\,\ell_P} \left[1-
    \frac{m\,\ell_P}{\sqrt{2+(m\,\ell_P)^2}}\right].
\ee
Physically, the non-commutativity or equivalently the boundness of the space of momenta regularizes the
ultra-violet divergences of the classical propagator.

At large distances, the propagator
$G(x;x')$ should coincide with $G_{cl}(x;x')$.
To see this is indeed the case, let us note
that $G(x;x')$ is in fact a function
$F(r/\ell_P,m\,\ell_P)$ from the expression (\ref{GX}).
Then, it becomes clear that
the large distance limit is defined by the condition
$\ell_P \rightarrow 0$ with $m$ and $x$ fixed.
As a result, we have
\be
    G(x;x') \sim G_{cl}(x;x') \,.
\ee
Therefore, we recover
the classical behavior of the propagator.

\section{Fuzzy space formulation of $C(SU(2))^*$}

This section is devoted to show that $C(SU(2))^*$ can be described
as a fuzzy space. This is done, in a first part, by defining a
notion of Fourier transform on the non-commutative algebra of
distributions $C(SU(2))^*$\,. Their image by this Fourier
transform is an algebra of matrices. In a second part, we
concentrate on symmetry aspects and show how the action of
$\mathcal{D}(SU(2))$ is expressed in the fuzzy formulation.

\subsection{Fourier transform of $C(SU(2))^*$}

Let us start by recalling that $C(SU(2))^*$ is the convolution
algebra of distributions on $SU(2)\,$. The Fourier transform on
$C(SU(2))^*$ can be induced by that on $C(SU(2))$\,. Given a
compact Lie group $G$\,, the Fourier transform on $C(G)$ is
defined using harmonic analysis on the group: the Fourier
transform of a given function in $C(G)$ is  its decomposition on
the  unitary irreducible representations (UIR) of the group $G$\,.
In the case of $SU(2)$\,, UIRs are characterized by a half-integer
spin $j$ and their basis vectors are labelled by the magnetic
number $m\in [-j,+j]$\,: \be\label{representationsofsu2}
    \mathcal J^2\vert j,m\ra=j(j+1)\vert j,m\ra\,,
    \qquad
    \mathcal J_3\vert j,m\ra=m \vert j,m\ra\,.
\ee
It is useful to introduce the Wigner $D$-matrices
$D^{j}_{mm'}(u)\equiv\la j,m\vert\, u\,\vert j,m'\ra$ as matrix elements of
representations.

The $SU(2)$ Fourier transform is then defined as the following map:
\ba\label{Fourier}
    {\cal F} \, : \, C(SU(2)) & \longrightarrow & \text{Mat}(\mathbb C)
     \equiv \bigoplus_{n\in \mathbb{N}} \text{Mat}_{n\times n}(\mathbb C) \nn
    f & \longmapsto & {\cal F}[f]
    \equiv\bigoplus_{2j+1\in \mathbb{N}}\int d\mu(u)\,D^j(u^{-1})\,f(u)\,.
\ea
The inverse map $\mathcal{F}^ {-1}:\text{Mat}(\mathbb C) \rightarrow C(SU(2))$
is given by
\be\label{inverseFourier}
    \mathcal{F}^{-1}[M](u)\equiv\text{Tr}\left(D(u)\,M\right)\,,
    \qquad
    \forall\,M=\bigoplus_{n\in\mathbb{N}} M_{(n)}\in \text{Mat}(\mathbb{C})\,,
\ee
where $M_{(n)}\in\text{Mat}_{n\times n}(\mathbb{C})$
and the trace $\text{Tr}$ in $\text{Mat}(\mathbb C)$ is defined as
\be
    \text{Tr}\,M\equiv\sum_{n\in\mathbb{N}} n\,\text{tr}\,M_{(n)}\,.
\ee
We have also introduced the notation $D$ for the $\text{Mat}(\mathbb C)$-valued
$SU(2)$ functions defined by $D(u)=\bigoplus_j D^j(u)$\,.

The Fourier transform $\mathcal{F}[\phi]$ of
a distribution $\phi \in C(SU(2))^*$ is a linear map on
$\text{Mat}(\mathbb C)$ defined by the relations
\be\label{fourier dist}
    \text{Tr}\left(M^\dagger\,\mathcal{F}[\phi]\right)
    \equiv \la \mathcal{F}^{-1}[M]\,,\,\phi \ra\,,
    \qquad \forall\,M\in \text{Mat}(\mathbb{C})\,.
\ee
Using the equation (\ref{inverseFourier}),
we get a more explicit and simple expression of $\mathcal{F}[\phi]$\,:
\be
    \mathcal{F}[\phi]=\la D^t\,,\,\phi\ra\,.
\ee When $\phi$ belongs to $C(SU(2))$\,, this formula coincides
with (\ref{Fourier}).

Finally, we have the ingredients to write explicitly the Fourier
transform of $C(SU(2))^*$\,: to any distribution $\phi$, we
associate an element $\widehat\Phi$ of $\text{Mat}(\mathbb C)\,$
as \be
    \label{matrix}
    \widehat\Phi\equiv\mathcal{F}[\phi]=\la D^t \,,\,\phi\ra\,.
\ee This map is invertible and its inverse reads \be
    \phi=\text{Tr}\,\big(\widehat\Phi\,D\big)\,.
\ee
The Fourier map defines a morphism and a co-morphism. As a result,
the space $\text{Mat}(\mathbb C)$ inherits a Hopf-algebra structure.
Of particular interest is the algebra structure of $\text{Mat}(\mathbb C)$
inherited from the convolution product in $C(SU(2))^*\,$ which is formally defined as follows:
\be
    \widehat\Phi_1 \star \widehat\Phi_2
    \equiv \la D^t \, , \, \phi_1 \star \phi_2 \ra
    =\la \Delta_{C(SU(2))}(D^t) \, , \, \phi_1 \otimes \phi_2 \ra \,.
\ee
In fact, this product is the standard matrix product in $\text{Mat}(\mathbb C)$\,:
\be
    \widehat\Phi_1 \star \widehat\Phi_2 = \widehat\Phi_1\,\widehat\Phi_2\,.
\ee
This is a consequence of $\Delta_{C(SU(2))}(D^{j}_{mm'})=\sum_n D^{j}_{mn} \otimes D^{j}_{nm'}\,$.
Therefore the algebra $C(SU(2))^*$ is isomorphic to the algebra of matrices
$\text{Mat}(\mathbb{C})$ and in the following we will omit
the notation $\star$ for the matrix product.

As a final remark about the Fourier transform, let us write the
 scalar product between two distributions $\phi_1$ and $\phi_2$ in terms of
their Fourier modes:
\be
    \la \phi_1\,,\,\phi_2\ra=
    \text{Tr}\,\big({\widehat{\Phi}_1}^\dagger\,\widehat{\Phi}_2\big)\,.
\ee
The obtained scalar product is the usual one defined in the fuzzy sphere when
restricted to one single space (or fuzzy sphere) $\text{Mat}_{n\times n}(\mathbb C)$.

\subsection{$\mathcal{D}(SU(2))$ symmetry in the Fuzzy space}

From the very beginning, we know that $\mathcal{D}(SU(2))$ plays a
crucial role in the construction of $C(SU(2))^*$\,:
$\mathcal{D}(SU(2))$ can be viewed as the isometry algebra of the
non-commutative space. We have already expressed the action of
$\mathcal{D}(SU(2))$ on $C(SU(2))^*$ in eq.(\ref{actionofDSU2}).
It is easy to export this action to the matrix algebra
$\text{Mat}(\mathbb C)$ as follows: \be\label{generalsymmetry}
    (f\otimes u) \rhd \widehat\Phi 
    \equiv {\cal F}[(f,u)\rhd \phi]
    = \la \text{ad}_{u^{-1}}(\overline{f}\,D^t)\,,\,\phi\ra\,,
\ee where $(f\otimes u) \in \mathcal{D}(SU(2))\,$, $\phi$ is a
distribution in $C(SU(2))^*$
 and $\widehat\Phi$ its Fourier transform.
In this section, we want to express this action explicitly. For
purposes of clarity, we will study separately actions of rotations
in $SU(2)\subset\mathcal{D}(SU(2))$ and those of translation-like
elements in $C(SU(2))\subset\mathcal{D}(SU(2))$\,.

\subsubsection*{Action of $SU(2)$}

We know that rotations act by conjugacy on
$C(SU(2))^*$ (see eq.(\ref{actionofDSU2})).
As a consequence, the action of $u\in SU(2)$ on a matrix family $\widehat\Phi$ simply reads
\be
    \label{rotation_fuzzy}
    (1\otimes u)\rhd\widehat\Phi \,=\,
    D(u)\,\widehat\Phi\,D(u^{-1})\,,
\ee As expected, rotations leaves the fuzzy spheres
$\text{Mat}_{n\times n}(\mathbb C)$ invariant. We can deduce the
action of an infinitesimal rotation by $\mathcal{J}_a\in
\mathfrak{su}(2)$ on $\widehat\Phi\,$: \be
    (1\otimes \mathcal{J}_a)\rhd\widehat \Phi=
    \big[\,D(\mathcal{J}_a)\,,\,\widehat\Phi\,\big]\,.
\ee

\subsubsection*{Action of $C(SU(2))$}

Translation-like elements $(f\otimes e)$ act by multiplication of
$f$ in the space $C(SU(2))^*$\,. This action induces the following
one on the space of matrices: \ba
    (f\otimes e)\rhd \widehat\Phi
    \e  \la \overline{f}\, D^t\,,\,\phi\ra
    =\text{Tr}\big(\,\mathcal{F}[\,\overline{f}\,D\,]^\dagger\;\widehat\Phi\,\big)
    \nn
    \e  \int d\mu(u)\,f(u)\,D(u^{-1})\,\text{Tr}\big(\,D(u)\,\widehat\Phi\,\big)\,.
\ea To have a more intuitive idea, let us look at infinitesimal
translations generated by the momentum  $\mathscr{P}_a=(P_a,e)$
defined by \be
    P_a(u)={{\ell_P}}^{-1}\,\sin(\theta/2)
    ={{\ell_P}}^{-1}\,i\,\text{tr}^{\frac{1}{2}}(\mathcal J_a\,u)\,.
\ee
The notation $\text{tr}^{\frac{1}{2}}$ holds for the trace in the fundamental $SU(2)$
representation.
The calculation of $\mathop{\mathscr{P}_a\rhd\widehat\Phi}$ mimics the previous one and
we get after a straightforward calculation that \ba\label{momentum
action}
    (\mathscr{P}_a\rhd\widehat\Phi)^j_{st}
    &\!\!=\!\!&
    \frac{i}{{\ell_P}} \sum_k (-1)^{m-n}\,\widehat\Phi^j_{mn}\, D^{1/2}_{pq}(\mathcal{J}_a)
    \left( \begin{array}{ccc} 1/2 & j & k \\ q & s & -m \end{array} \right)
    \left( \begin{array}{ccc} 1/2 & j & k \\ p & t & -n \end{array} \right)
    \nn
    &\!\!=\!\!&
    \frac{i\,D^{1/2}_{pq}(\mathcal{J}_a)}{{\ell_P}\, (2j+1)}\,
    \Big(\sqrt{(j+1+2qs)(j+1+2tp)} \; \widehat\Phi^{j+1/2}_{q+s \, p+t} \nn
    &&\qquad\qquad\qquad
     +\,(-1)^{q-p}\sqrt{(j-2qs)(j-2pt)} \; \widehat\Phi^{j-1/2}_{q+s \, p+t} \Big)\,,
\ea where we have used the notations of the book \cite{Edmonds}
for the Clebsh-Gordan coefficients. From the above formula, we see
that infinitesimal translations move points of the sphere
$\text{Mat}_{(2j+1)\times (2j+1)}(\mathbb C)$ into points of the
nearest neighbor spheres. This is exactly what one would expect.


\section{Relation between $\text{Mat}(\mathbb C)$ and $C_{{\ell_P}}(\mathbb E^3)$}

This section is devoted to establish a general correspondence
between the space of matrices $\text{Mat}(\mathbb C)$ and the
deformed algebra of functions $C_{{\ell_P}}(\mathbb E^3)\,$. This
correspondence   is  important  in order to understand for
instance the undeformed limit of the space of matrices.

\subsection{The general relation}

For that purpose, it is necessary to decompose the matrix $\widehat\Phi$, associated to a distribution $\phi$,
into a multiplet $\widehat\Phi_+\oplus\widehat\Phi_-\oplus\widehat\Phi_0$ of matrices associated to the multiplet
$\phi_+\oplus\phi_-\oplus\phi_0$.
Each component $\widehat\Phi_\varepsilon$
is the matrix representation of $\phi_\varepsilon$ and can be obtained from the matrix $\widehat\Phi$
as follows:
\ba\label{Mdecomp}
\widehat\Phi_{\pm}^j = \int d\mu(g) \, \text{Tr}(\widehat\Phi D(g)) \, D^j(g) \, I_\pm(g) \;\;\;\;
\text{and} \;\;\;
\widehat\Phi = \widehat\Phi_++\widehat\Phi_-+\widehat\Phi_0
\ea
where $I_\pm(g)= \Theta(\pm(\pi-\theta))$ is characteristic functions written in terms of
the theta function: $\Theta(x)=1$ if $x>0$ otherwise 0.

Now, we can establish a link between each component $\widehat\Phi_\varepsilon$ and the functions
$\Phi_\varepsilon(x)$. We start with the cases $\varepsilon=\pm$.
Using  the relations  (\ref{relation Phiphi}) and (\ref{matrix}), we obtain
the mapping from $\widehat{\Phi}\in\text{Mat}(\mathbb C)$ to $\Phi_\pm \in C_{{\ell_P}}(\mathbb E^3)$:
\ba\label{Mat to Fun}
    \Phi_\pm(x) \, = \, \la e^{iP\cdot x}\chi_\pm\,,\,\phi_\pm\ra
    = \text{Tr}\left(\mathcal{F}^\dagger[e^{iP\cdot x}I_\pm]\,
   \mathcal{F}[\phi_\pm]\right)
    = \text{Tr}\,\big(K_\pm^\dagger(x) \,\widehat\Phi_\pm\big)
\ea
where $e^{iP\cdot x}$ is the plane wave viewed as a function on $SU(2)$.
This mapping is by construction a morphism.
In the last identity, we introduced the notation $K_\pm$
for the $\text{Mat}(\mathbb C)$-valued function on $\mathbb E^3$
explicitly defined by $\mathop{K_\pm(x)\equiv \mathcal{F}[e^{iP\cdot x}I_\pm]}$\,.
The functions $K_\pm$ can be interpreted as the components of an element
$K \in C_{{\ell_P}}(\mathbb{E}^3) \otimes \text{Mat}(\mathbb{C})$ given by:
\be\label{def of K}
    K(x)= K_+(x) \xi_+ + K_-(x) \xi_- =
    \la w_x\,,\,D\ra = \int d\mu(u) \, D(u) \, \overline{w_x(u)}.
\ee It remains to obtain the relation between $\widehat{\Phi}_0$
and $\Phi_0(x)$. To do so, we follow the same idea as previously.
We start by writing the relation between $\phi_0$ and $\Phi_0(x)$
making use of the momentum $\vec P_0$ and then we expand $\phi_0$
into its Fourier modes $\widehat{\Phi}_0$ to obtain the relation:
\be\label{relPhi0phi0} \Phi_0(x) \, = \, \la e^{iP_0\cdot
x}\,,\,\phi_0\ra \, = \, \text{Tr}(K_{\nu_0}^\dagger(x) \,
\widehat\Phi_0) \ee where the matrix valued function $K_{\nu_0}$
depends on a vicinity $\nu_0 \subset SU(2)$ of $U_0$ where $P_0$
is well-defined according to the integral: \be K_{\nu_0}(x) \, =
\, \int d\mu(u) \, D(u) \, \exp(-i{P}_0(u) \cdot x) \,
I_{\nu_0}(u) \ee where $I_{\nu_0}$ is the characteristic function
on $\nu_0$. It is clear that $K_{\nu_0}$ depends on the choice of
$\nu_0$ but the relation (\ref{relPhi0phi0}) is independent of
that choice.

The relations (\ref{Mat to Fun}) and (\ref{relPhi0phi0}) are
invertible and their inverse can be obtained using inverse Fourier
transform. Thus, one uniquely associates to any element of
$C_{{\ell_P}}(\mathbb E^3)$\,, a family of matrices in
$\text{Mat}(\mathbb{C})$.

One can interpret the function $\Phi_{\varepsilon}(x)$
as a kind of continuation to the whole Euclidean space
of a discrete function ${\widehat\Phi_{\varepsilon}}{}^j_{mn}$
which is a priori defined only on a infinite but
denumerable set of points:
there are obviously $(2j+1)^2$ points on each fuzzy sphere of dimension $2j+1$\,.
Given any $x\in \mathbb E^3$\,,
each matrix element ${\widehat\Phi_{\varepsilon}}{}^j_{mn}$
contributes to the definition of $\Phi_{\varepsilon}(x)$ with a complex weight $\overline{K^j_{\pm nm}(x)}$
or $\overline{K^j_{\nu_0 nm}(x)}$\,.

We conclude  this section with some properties of the elements $K_\pm$ and $K_{\nu_0}$.
\begin{enumerate}
    \item $K_\pm$ are normalized elements in both $C_{{\ell_P}}(\mathbb{E}^3)$
    and $\text{Mat}(\mathbb{C})$\, in the sense that:
    \be\label{normalization K}
    \int \frac{d^3x}{(2\pi)^3\,v_{{\ell_P}}}\,K_\pm(x)=\mathbb{I}_\pm\,,
    \qquad
    \text{Tr}\left(K(x)\right)=\xi_+\,,
    \ee
    where $\mathbb{I}_\pm=\bigoplus_{n\in\mathbb{N}}(\pm 1)^{n+1}\mathbb{I}_{(n)}$
    with $\mathbb{I}_{(n)}$  the $n\times n$ identity matrix.
    \item The previous properties can be extended to the matrix-valued function $K_{\nu_0}$.
    Assuming that $P_0$ does not vanish in $\nu_0$ and $\nu_0$ does not contain the identity,
    then we have:
    \be
    \int \frac{d^3x}{(2\pi)^3\,v_{{\ell_P}}}\,K_{\nu_0}(x)=0,
    \qquad
    \text{Tr}\left(K_{\nu_0}(x)\right)=0 \,.
    \ee
    \item The matrix elements of the functions $K$
    form clearly a sub-algebra of $C_{{\ell_P}}(\mathbb E^3)$
    and their product is given by
    \be
    \left( K^j_{mn}\star K^{j'}_{m'n'}\right)\!(x)
    =\frac{\delta_{jj'}}{2j+1}\,\delta_{m'n}\,K^j_{mn'}(x)\,.
    \ee These relations are consequences of the fact that the matrix
    elements of $SU(2)$ UIRs are orthonormal when viewed as functions
    on $SU(2)$.
\end{enumerate}

\subsection{The coordinates on the Fuzzy space}

In this section we determine  and study the coordinate functions
$x^a \,\xi_+\in C_{{\ell_P}}(\mathbb{E}^3)$ in the fuzzy space representation $\text{Mat}(\mathbb C)$\,.
To construct the coordinate functions, we use the plane waves $w_u(x)$
which can be viewed as a coordinate generating function.

Let us start by finding the matrix representation of plane waves.
Plane waves have been defined previously as the image in $C_{{\ell_P}}(\mathbb{E}^3)$
of delta distributions $\delta_u$. As a result, the matrix representation of a plane wave reads:

\be
    \widehat{w}_{u}=\la D^t\,,\,\delta_u\ra=D(u^{-1})\,.
\ee
Using the parametrization (\ref{parametrisation of su2}) of $SU(2)$
and doing the following expansion in the angle $\theta$ of each side of the above equation:
\ba
    &&
    \widehat{w}_{u}=
    \mathbb{I}+2i\,\theta\,n^a\,\Big[{{\ell_P}}^{-1}\,\widehat{x_a}\Big]
    -2\,\theta^2\,n^a n^b
    \left[{{\ell_P}}^{-2}\,\widehat{x_a\,x_b}\right]
    +\mathcal{O}(\theta^3)\,,
    \nn
    &&
    D(u^{-1})=
    \mathbb{I}+i\,\theta\,n^a\,\Big[D(\mathcal{J}_a)\Big]
    -\frac{\theta^2}{2}\,n^a n^b
    \left[\frac{1}{2}\,D(\mathcal{J}_a\,\mathcal{J}_b+\mathcal{J}_b\,\mathcal{J}_a)\right]
    +\mathcal{O}(\theta^3)\,.
\ea
one finds immediately (at the first order in $\theta$)
the matrix representations $\widehat{x}_a$ of the coordinate functions $x_a\xi_+\in C_{{\ell_P}}(\mathbb{E}^3)$
which are just the matrix representation of the $\mathfrak{su}(2)$ generators $\mathcal{J}_a$\,:
\be
    \widehat{x}_a=2\,{\ell_P}\,D(\mathcal{J}_a)\,.
\ee As expected, the coordinates are non-commutative functions
which satisfy the $\mathfrak{su}(2)$ Lie-algebra relation. This
relation  holds as it should  in the $C_{{\ell_P}}(\mathbb E^3)$
representation and read the commutator (\ref{coord commuator})
between coordinate functions with $\star$-product.
Furthermore, at the second order in $\theta$\,, 
one obtains more than the commutator between coordinates
but the product of coordinates (\ref{coord algebra}).

We recover the well-known fact that coordinates functions on fuzzy
spheres are obtained from representations of the
$\mathfrak{su}(2)$ Lie algebra generators. Note that, in our
construction, this fact is not put by hand but is a consequence of
some more fundamental principles. The geometrical consequences of
this result are immediate: the spectrum of coordinates is discrete
and can be obtained from $\mathfrak{su}(2)$ representations
theory. In particular, the radius (squared) matrix $\widehat{r_\star}$ with
$r^2_\star\,\xi_+=x_a\xi_+\star x^a\xi_+$ is proportional to the identity
matrix on sphere $\text{Mat}_{n\times n}(\mathbb C)$ and its
values are given by the $\mathfrak{su}(2)$ Casimir evaluated in
the representation $j=(n-1)/2\,$, i.e.
$\widehat{r_\star}=2\sqrt{j(j+1)}\,{\ell_P}\,\mathbb{I}_{n\times
n}\,$. Thus the algebra of matrices $\text{Mat}(\mathbb{C})$ can
be thought as the dual of concentric fuzzy spheres or a fuzzy
onion.

\subsection{Examples}

To have a more intuitive idea of the non-commutative space,
let us now give some simple examples of functions and their equivalent formulations
in $C(SU(2))^*$, $C_{{\ell_P}}(\mathbb{E}^3)$ and $\text{Mat}(\mathbb C)$.

\subsubsection*{Constant function}

Constant functions are the simplest examples which correspond in
$C(SU(2))^*$ to the elements $\phi=c_+\,\delta_e + c_-\,\delta_{e_A} $ where
$c_+,c_-$ are constant.
We deduce immediately its $C_{{\ell_P}}(\mathbb E^3)$ and $\text{Mat}(\mathbb C)$ representations
given by
\be
    \widehat\Phi^j=c_+ \mathbb I^j  \, + \, (-1)^j \, c_- \mathbb I^j ,\qquad
    \Phi_\pm(x) = c_\pm\,.
\ee

One-dimensional functions are also simple but less trivial examples.
Their set is defined as the kernel of two out of the three derivative operators $\partial_a$.
They form a commutative algebra which is not, nonetheless, the point-wise product.
These functions and some applications have been studied in \cite{Noui2}.

\subsubsection*{Radial function}

A classical radial function depends only on the radius
$r=\sqrt{x^a\,x_a}\,$ and  is obviously invariant under rotations.
In the deformed context, a radial function is also defined to be
invariant under rotations. As a consequence, in the $C(SU(2))^*$
representation, a radial function corresponds to a distribution
$\phi$ on the conjugacy classes of $SU(2)$\,. In the sequel, we
consider only the cases where $\phi$ is a function for simplicity.
One immediately obtains that the matrices $\widehat \Phi_\pm^j$
are proportional to the identity matrix: \be
    \widehat\Phi^j_\pm
    ={\lambda^{(j)}_\pm}\,\mathbb{I}_{(2j+1)}\,.
\ee
In the $C_{{\ell_P}}(\mathbb E^3)$ representation, $\Phi_\pm$ can be
expressed as a series:
\be\label{Mat to Fun r}
    \Phi_\pm(r)=
    8\pi\sum_{2j+1\in\mathbb{N}}
    {\ell_P}\,{r_j}^2\,\lambda^{(j)}_\pm \,R_\pm^j(r)\,,
\ee
with the discrete radius $\mathop{r_j\equiv (2j+1)\,{\ell_P}}$ and
the radial modes $R_\pm^j(r)$ are given by
\be
    R_\pm^j(r) \equiv \frac{1}{2(2\pi)^3}\frac{1}{(2j+1)\,v_{{\ell_P}}} \text{tr}(K_\pm^j(x))\,.
\ee
Let us underline some properties of the functions $R_\pm^j$\,:
\begin{enumerate}
    \item A more explicit integral formula for the functions $R_\pm^j$\, can be obtained
    as follows:
    \ba
        R_\pm^j(r)
        &\!\!=\!\!& \frac{1}{2(2\pi)^3}
        \frac{1}{(2j+1)\,v_{{\ell_P}}}\int d\mu(u)\,{C}^j(u)\,w_{u^{-1}}(r) \, I_\pm(u)
        \\
        &\!\!=\!\!&
        \frac{(\pm1)^{2j}}{4\pi\,r\,r_j}
        \left[
        {\ell_P}^{-1}
        \int_{0}^{\pi}\frac{d\theta}{2\pi}\,
        \sin\!\left(\frac{r_j}{2\,{\ell_P}}\,\theta\right)
        \sin\!\left(\frac{r}{{\ell_P}}\,\sin(\theta/2)\right)
        \right],\nonumber
    \ea
    where ${C}^j(u)=\text{tr}(D^{j}(u))$ are the $SU(2)$ characters.
    In fact, when $j$ is an integer $R_\pm^j$ can be expressed in terms of the Bessel function $J_n$
    as follows:
\be
R^j_\pm (r) \; = \; \frac{1}{4\pi\,r\,r_j}
        \left[\frac{1}{2\,{\ell_P}}\,\text{J}_{r_j/{\ell_P}}
        (r/{\ell_P})\right]\,.
\ee
    This is not true when $j$ is a half-integer.
    \item The functions $R^j$ are normalized when viewed as elements of $\text{Mat}(\mathbb C)$
    as well as when viewed as elements of $C_{{\ell_P}}(\mathbb E^3)$\,:
    \be\label{normR}
        4\pi\int dr\,r^2\,R_\pm^j(r)= \frac{(\pm 1)^{2j}}{2} \, ,\quad
        4\pi\sum_{2j+1\in\mathbb{N}} (2\,{\ell_P})\,r_j^2\,R_\pm^j(r) = \frac{1\pm 1}{2}\, .
    \ee
    This is a consequence of the eq.(\ref{normalization K}).
\end{enumerate}

There is a nice interpretation of the decomposition formula (\ref{Mat to Fun r}) as the
Riemman sum of the classical integral:
\be
    \Phi(r) =  \int_0^\infty d\rho \, \Phi(\rho) \, \delta(r-\rho) \,.
\ee
For such an interpretation to be true, it is necessary to compute the classical limit
of the radius modes $R_\pm^j$\,. First of all, we defined the classical limit by
$\mathop{{\ell_P}\to0}$ but $r$ and $r_j$ finite. We note that, in the integral formula of $R_\pm^j$, the
two sine functions in the integrand oscillate very fast
and the integral vanishes unless $r=r_j$\,.
Since $R_\pm^j(r)$ is normalized to $\frac{(\pm 1)^{2j}}{2}$ according to the formula (\ref{normR}),
the radial modes $R_\pm^j(r)$ tend to
the delta distribution $\mathop{\delta(r-r_j)}$, up to a global factor, in the undeformed limit:
        \be
        R_\pm^j(r)
        \underset{{\ell_P}\to 0}{\longrightarrow}\,
        \frac{(\pm 1)^{2j}}{8\pi\,r^2}\,\delta(r-r_j)\,.
        \label{int_R}
        \ee
        This formula shows that in the undeformed limit,
        the value of radial function $\Phi_\pm$ at $r=r_j$ is
        given by the matrix element $\lambda^{(j)}_\pm$\,:
        $\Phi_\pm(r_j)=(\pm 1)^{2j}\lambda^{(j)}_\pm\,$.

\subsubsection*{Delta function}

Other important examples are the delta distributions for the $\star$-product.
As in the classical case, the delta-distribution localized at the point $x$ is, in the $C(SU(2))^*$ formulation,
the plane wave $w_x(u)$. Thus, as one could expect, the delta-distributions localized at $x$ are related to the delta
distribution localized at the origin by a translation, namely $w_x = T_{-x}\rhd w_0$. For that reason we
concentrate first on the delta distributions localized at the origin which is in fact the constant function $1$ in the
$C(SU(2))^*$ formulation.

In the fuzzy space formulation, the delta distribution at the origin is given by the matrix
$\widehat{\delta}_{\star 0}$ whose elements are
\be
\widehat{\delta}_{\star 0}^j \; = \; \delta_{j,0} \;.
\ee
As the result, the matrix is completely localized at the fuzzy origin.

In order to get its $C_{\ell_P}(\mathbb E^3)$ formulation, one has
to decompose the constant function $1$ into its three components
$1_+\oplus1_-\oplus 1_0$. It is immediate to see that
$1_\pm(g)=I_\pm(g)$ and $1_0=0$ where $I_\pm$ have been introduced
in (\ref{Mdecomp}). Note that there is no zero component for the delta
distribution. 
Furthermore, the functions $\delta_{\star 0\pm} \in
C_{\ell_P}(\mathbb E^3)$ are radial. One can easily show that they
are equal and given by: \be \delta_{\star 0\pm}(r) \; = \;
\frac{1}{\pi r} \int_{0}^{\pi}\!\!d\theta \,
\sin(\frac{\theta}{2}) \,
\sin(\frac{r}{\ell_P}\sin(\frac{\theta}{2})) \; = \;
\frac{1}{2}R^0(r) \ee where $R_\pm^j$ have been introduced in the
previous examples. The delta-distribution satisfies the following
expected relation: \ba\label{delta_star}
    \xi_+^t\int \frac{d^3x}{(2\pi)^3 v_{\ell_P}}\,
    (\Phi \star \delta_{\star 0})(x)
    = \, \Phi_+(0)\, + \, \Phi_-(0) ,
    \qquad \forall\,\Phi_\varepsilon \in C_{{\ell_P}}(\mathbb E^3) \,.
\ea
Since $w_y=T_{-y}\,w_0$\, we find that the delta distributions localized at any point
$y$ satisfy
\be
    \delta_{\star y\pm}(x)=(\mathscr{T}_y\rhd \delta_{\star0\pm})(x)=\delta_{\star 0\pm}(x-y)\,.
\ee

The main difference with the classical case is that $\delta_{\star
0}$ is spread around the origin. It is nevertheless the most
localized function at the origin: we cannot localized the origin
of the fuzzy space with a better precision due to the
non-commutativity of the coordinates. Besides,
$\delta^\star_{0\pm}$ tends to the classical delta distribution on
$\mathbb E^3$ in the undeformed limit because $R^0$ does as we
have seen it previously.

When considering in (\ref{delta_star}) $\Phi(x)=x_a\xi_+$  we recover
the fact that $\delta^\star_y$ is localized at the point $y$\,.
Nevertheless, as for the function $\delta^\star_0$\,,
$\delta^\star_y$ is spread. This fact can be illustrated computing
the previous integral with $\Phi(x)=x_a\xi_+\star x_b\xi_+$\,:
 \be
    \int \frac{d^3x}{(2\pi)^3 v_{\ell_P}}\, x_a \xi_+\star x_b \xi_+ \star \delta^\star_y(x)
    =y_a \xi_+ \star y_b \xi_+ \neq y_a\,y_b \,\xi_+ \,.
\ee
As a result, the integral is different
from its undeformed analog.

\subsection{Maximally localized state}

As shown in the previous lines, $\delta^\star_0$ distributions are interpreted as the most
localized functions at the origin in $C_{{\ell_P}}(\mathbb E^3)$.
In this  section we determine
the functions which are the most localized around a classical point.

To do so, we work in the fuzzy space representation. We define the
most localized function at a given classical point
$x\in\mathbb{E}^3$ as the one which corresponds to the minimum of
the operator $(\widehat{x_a}-x_a)(\widehat{x^a}-x^a)$, viewed as a
family of matrices.

The minimum of the previous operator is realized on the state
$|j\,\vec{e_x}\ra$ where $\vec{e_x}$ is the unit vector in the
direction of $\vec x$ and   the state is such that: first, it is
an eigenvector of $\vec{\mathcal{J}}\cdot \vec{e_x}$ with
eigenvalues $j$; second, it is an eigenvector of $\mathcal{J}^2$
with eigenvalue $j(j+1)$; third, the eigenvalues are such that:
\be
    2j+\frac{1}{2}=\frac{r_j}{{\ell_P}}-\frac{1}{2}
    =\left\lceil \frac{r}{{\ell_P}}-\frac{1}{2}\right\rceil.
\ee
where $\lceil x \rceil$ is the Ceiling function of $x$ namely the smallest integer 
not less than $x$\,.
The minimum of the previous operator is thus given by
\be
    \min\!\left[(\widehat{x}-x)^2\right]
    ={{\ell_P}}^2
    \left\{ (\lceil a \rceil-a)^2+2a \right\},
    \qquad
    a=\frac{r}{{\ell_P}}-\frac{1}{2}\,.
\ee

Notice that as a function of $r$ this minimum has local minima for
$a=\lceil a \rceil$, i.e. for integer radius $r=r_{j'}$ ($2j'+1$
being an integer).

Therefore, the projector in $\text{Mat}(\mathbb{C})$ onto the
state $|j\, \vec\omega\ra$ is the function which is the most
localized  around a classical point $r_j \vec{\omega}$. Its
associated $C(SU(2))^*$ element is obtained via Fourier transform
and is explicitly given by: \be
    \phi_{j,\vec\omega}\big(u(\theta,\vec n)\big)
    =\la j\,\vec\omega|\,D^{j}\big(u(\theta,\vec n)\big)\,|j\,\vec\omega\ra
    =\left(\cos\frac{\theta}2-
    i\,\vec n\cdot\vec\omega\,\sin\frac{\theta}2\right)^{2j}\,.
\ee It is interesting to note that the classical limit of such a
function, obtained by considering the limit  $\mathop{{\ell_P}\to0}$
with $r_j$ fixed is given by \be
    \lim_{{\ell_P}\to0}\phi_{j,\vec\omega}(u)
    =\exp\left(-i\,r_j\,\omega^a\,P_a(u)\right)\;.
\ee
This result agrees with the fact that the limit of the $C_{{\ell_P}}(\mathbb{E}^3)$ representation
$\Phi_{r_j,\vec\omega}(x)$ of $\phi_{j,\vec\omega}/v_{{\ell_P}}$
is the usual delta distribution:
\be
 \lim_{{\ell_P}\to0}\Phi_{r_j,\vec\omega}(x)=(2\pi)^3 \, \delta^3(\vec x-r_j\,\vec\omega)\,.
 \ee

\section{Conclusion}

In this paper, starting from the deformation of the Euclidean
group we constructed a non-commutative space carrying tha action
of $ISU(2)$.

This non-commutative space has been described, as usual in non-commutative geometry,
by its algebra of functions which has been shown to be the algebra $C(SU(2))^*$ of $SU(2)$
distributions endowed with the convolution product.
We found two additional  representations
of this algebra.
The first is given by $C_{{\ell_P}}(\mathbb E^3)$.
    Its elements  are given by  three functions on $\mathbb E^3$,
    two of them having a spectrum strictly bounded by $\ell_P^{-1}$, 
    the last one having its spectrum
    on the sphere of radius $\ell_P^{-1}$.
    This representation
    makes a bridge between the non-commutative space and 
    the standard classical manifold $\mathbb E^3$.
    The important point is that the product of the algebra 
    $C_{{\ell_P}}(\mathbb E^3)$ is no-longer the point-wise product but
    a $\star$-deformation of it with ${\ell_P}$ as a deformation parameter.
    The second representation is  obtained by means of matrices.
    To get this formulation 
    we introduced a Fourier transform on the algebra $C(SU(2))^*$ 
    whose Fourier space
    is the algebra of complex matrices $\text{Mat}(\mathbb C)$. 
    This shows that the quantum space we constructed
    is in fact discrete or fuzzy, as expected from the boundedness of the 
    momenta space. 

Finally, we showed the correspondence between the continuous
formulation in terms of $C_{{\ell_P}}(\mathbb E^3)$ and the fuzzy
formulation in terms of $\text{Mat}(\mathbb C)$. This
correspondence gives some insights regarding   the geometry of the
fuzzy space and  its classical limit. Furthermore 
we illustrated our construction with 
examples. Some 
of these examples
 have been recently considered  in \cite{FM}.

This study is the starting point for a construction of a Quantum
Field Theory on the non-commutative space. A local action for a
scalar field has been proposed in Section 4. Similar actions have
been partially investigated in \cite{FL,Noui,Japanese} in the
momentum representation. We showed that the appearance of a
discrete degree of freedom is unavoidable in this context leading
to a multiplet of scalar fields. It would be interesting to
consider the formulation of such QFTs in the fuzzy space
representation and its links to matrix models for instance.

Our construction is very general and opens the possibility to
extend our result to the Lorentzian sector and to the
non-vanishing cosmological constant cases, namely de Sitter and
anti de Sitter space times. Indeed, in three dimensions, whatever
the sign of the cosmological constant and the signature of the
metric, quantum gravity is argued to have quantum doubles as
quantum isometry algebras \cite{BS}. In the Lorentzian sector, the
momentum space is no longer compact, therefore the fuzzy spacetime
is expected not to be fully discrete and the ultraviolet
divergences may not completely disappear. For de Sitter or anti de
Sitter spacetime, we expect the quantum deformations of the
classical isometry groups to play an important role. In these
cases, the momentum space is no longer a curved manifold but a
quantum group. It would be very interesting to look at the
associated quantum geometries which would be interpreted as de
Sitter or anti de Sitter quantum spaces.

\section*{Acknowledgments}
We would like to thank Renaud Parentani for discussions in
the early stages of the paper.
K.N. wants to thank E. Livine and A. Perez for discussions.
The work of K.N. was partially supported by the ANR
(BLAN06-3\_139436 LQG-2006). The work of J.M. was  partially supported
 by the EU FP6 Marie Curie
Research \& Training Network "UniverseNet" (MRTN-CT-2006-035863).

\appendix

\section{Schwinger's Oscillator representation of $SU(2)$ and the translations action}

In this appendix, we recall Schwinger's construction of the UIR of
$\mathfrak{su}(2)$ from a couple of commuting harmonic oscillators
defined by annihilation and creation operators $a_p$ and
$a_p^\dagger$ with $p \in \{+,-\}$ (see e.g.\cite{bied}). The
$\mathfrak{su}(2)$ generators $\mathcal{J}_a$ are given by:
\be\label{generatorsu2}
    \vec{\mathcal{J}} \, = \, \sum_{p,q} a^\dagger_p \, \frac{(\vec\sigma)_{pq}}2 \, a_q\,,
\ee where  $\vec\sigma\equiv 2\,D^{1/2}(\vec{\mathcal{J}}\,)$ are
the Pauli matrices. Furthermore, one can recover the unitary
irreducible representations of $\mathfrak{su}(2)$ from those of
the harmonic oscillators. The Fock space generated by $a_\pm$ and
$a_\pm^\dagger$ is spanned by given by the states $\vert
n_+\,;\,n_-\ra$ with \ba\label{irreposcillators}
    &a_+ \vert n_+\,;\,n_-\ra =
     \sqrt{n_+}\, | n_+ - 1\,;\,n_-\ra \,,
    &a_+^\dagger \vert n_+\,;\,n_-\ra =
     \sqrt{n_++1}\, | n_+ + 1\,;\,n_-\ra \,,
    \nn
    &a_- \vert n_+\,;\,n_-\ra =
     \sqrt{n_-}\, | n_+ \,;\,n_- -1\ra \,,
    &a_-^\dagger \vert n_+\,;\,n_-\ra =
     \sqrt{n_-+1}\, | n_+ \,;\,n_- +1\ra \,,\nn
\ea where $(n_+,n_-)$ is a couple of non-negative integers. If one
identifies the states $\vert n_+;n_-\ra$ and $\vert j,m \ra$ with
the relations $2j=n_++n_-$ and $2m=n_+-n_-$, then one recovers the
action of $\mathfrak{su}(2)$. As a result, the representation
(\ref{irreposcillators}) is not irreducible for
$\mathfrak{su}(2)$\,: it is the direct sum of the whole set of
finite dimensional representations of $\mathfrak{su}(2)$, each
representation appearing only once. Finally, let us recall the
link between occupation number $N=a_+^\dagger a_++a_-^\dagger a_-$
and the $\mathfrak{su}(2)$ spin $j$\,: \be N\vert j,m\ra \; = \;
\left(a_+^\dagger a_++a_-^\dagger a_- \right)\vert j,m\ra  \; = \;
2j\,\vert j,m\ra  \;. \ee

It is interesting to see how the infinitesimal translation can be
written in a more compact form within Schwinger's representation
of the $SU(2)$ UIRs: \be \label{expressioncanonique}
    \mathscr{P}_a\rhd\widehat\Phi
    =\frac{i\,(\sigma_a)_{mn}}{{\ell_P}\,(N+1)}
    \left(a_n \,\widehat\Phi \,a_m^\dagger - a_m^\dagger\,\widehat \Phi \,a_n\right),
\ee where $a_p$ and $a^\dagger_p$ are a couple of annihilation and
creation operators and $N=a^\dagger_p\,a_p$ is the occupation
number. Let us see that this is indeed the case and let us verify
that $\mathscr{P}_a$ satisfies the good properties of the
infinitesimal translation operator:
\begin{enumerate}
    \item The only non-vanishing matrix elements of
    $\delta_q \widehat\Phi\equiv iq^a\mathscr{P}_a\rhd \widehat\Phi$ are
    $\la j,s \vert \delta_q \widehat\Phi \vert j,t \ra $.
    Each term involving the operators $a_p$ and $a_p^\dagger$ can be computed using
    (\ref{irreposcillators})\,:
    \ba
        \la j,s \vert \, a_q \,\widehat\Phi\, a_p^\dagger\, \vert j,t \ra & = & \sqrt{(j+1+qs)(j+1+pt)}
        \; \widehat\Phi^{j+1/2}_{s+\frac{q}{2} \, t+\frac{p}{2}} \nonumber \\
        \la j,s \vert \, a_p^\dagger\, \widehat\Phi \,a_q\, \vert j,t \ra & = & \sqrt{(j+qt)(j+ps)} \;
        \widehat\Phi^{j-1/2}_{s-\frac{p}{2} \, t-\frac{q}{2}} \nonumber
    \ea
    If we use the fact that $(-1)^{m-n}D^{1/2}_{-m\, -n}(\mathcal{J}_a)=-D^{1/2}_{nm}(\mathcal{J}_a)$,
    then we recover the expression
    (\ref{momentum action}) for $\delta_q\widehat\Phi$ from (\ref{expressioncanonique}).

    \item The set of  hermitian matrices is stable under the action of translations $\delta_q$
    as one can trivially sees from the eq. (\ref{expressioncanonique}).

    \item The result of the translation of a coordinate variable $\widehat x_a$,
    represented
    by the matrices $2\,{\ell_P}\, D(\mathcal{J}_a)$
        in the fuzzy representation (see section 6.1 below), by a vector $q$ can be computed
    explicitly:
    \be
        \delta_q\, \widehat x_a \, = \, - \frac{2\,q^b}{N+1}\,
        D^{1/2}_{\frac{p}{2}\frac{q}{2}}(\mathcal{J}_b)\,
        D^{1/2}_{\frac{m}{2}\frac{n}{2}}(\mathcal{J}_a)
        \left(a_q a^\dagger_m a_n a_p^\dagger - a_p^\dagger a^\dagger_m a_n a_q\right).
    \ee
    As expected, one can show that $\delta_q\,\widehat x_a = q_a$\,.
    To do so, one has to use the standard commutation relations
    between creation and annihilation operators and the familiar property
    $D^{1/2}(\mathcal{J}_a\mathcal{J}_b)=1/2\,\epsilon_{abc}\,D^{1/2}(\mathcal{J}_c) + 1/4\,\delta_{ab}$\,.

    It is also interesting to remark that
    $\delta_q\,N = 4\,\left(q^a \,\mathcal{J}_a\right)/(N+1)$
    is the quantum analog of
    $\delta_q\,\widehat r = \left(q^a \,\widehat x_a\right)/\widehat r$\,.
    Indeed, $N$ and the radius $\widehat r$ are both scalar on each subspace
    $\text{Mat}_{d_j\times d_j}(\mathbb C)$
    and simply related by $N+1=\sqrt{4\,{\widehat r}^2 +1}$\,.
\end{enumerate}

\bibliographystyle{unsrt}

\begin{thebibliography}{10}




\bibitem{fuzzy}
J. Madore,
An introduction to Noncommutative Differential Geometry \& its applications,
Cambridge Univ. Press (2000);
A. Connes,
Noncommutative Geometry,
Academic Press (1994); R.J. Szabo,
Symmetry, gravity and noncommutativity,
Class. Quant. Grav. 23, R199-R242 (2006).

\bibitem{discrete}
A. Ashtekar and J. Lewandowski,
Background independent quantum gravity: a status report,
Class. Quant. Grav. 21, R23-R152 (2004);
A. Perez ,
Spin Foam models for quantum gravity,
Class. Quant. Grav. 20, R43-R104 (2003);
C. Rovelli,
Quantum Gravity,
Cambridge Univ. Press (2004);
T. Thiemann,
Introduction to modern canonical quantum general relativity,
Cambridge Univ. Press (2004).



\bibitem{sw}
M.R. Douglas and N.A. Nekrasov,
Noncommutative field theory,
Rev. Mod. Phys. 73, 977-1029 (2001).



\bibitem{chaichian}
M. Chaichian, P.P. Kulish, K. Nishijima and A. Tureanu,
On a Lorentz-invariant interpretation of noncommutative spacetime and its applications on
noncommutative QFT,
Phys. Lett. B604, 98-102 (2004).




\bibitem{antichaichian}
G. Fiore and J. Wess,
On "full" twisted Poincar\'e symmetry and QFT on Moyal-Weyl spaces,
ArXiv[hep-th/0701078];
E. Joung and J. Mourad,
QFT with Twisted Poincar\'e invariance and the Moyal product,
JHEP 0705:098 (2007).

\bibitem{DSR}
F. Girelli, E. Livine and D. Oriti,
Deformed special relativity as an effective flat limit of quantum gravity,
Nucl. Phys. B708, 411-433 (2004).


\bibitem{FL}
L. Freidel and E. Livine,
Ponzano-Regge model revisited III: Feynman diagrams and effective field theory,
Class. Quant. Grav. 23, 2021-2062 (2006);
L. Freidel and E. Livine,
Effective 3D quantum gravity and non-commutative quantum field theory,
Phys. Rev. Lett. 96: 221301 (2006).


\bibitem{Noui}
K. Noui,
Three dimensional Loop quantum gravity: particles and the quantum double,
J. Math. Phys. 47: 102501 (2006);
K. Noui,
Three dimensional Loop quantum gravity: towards a self-gravitating quantum field theory,
Class. Quant. Grav. 24, 329-360 (2007).


\bibitem{Bais}
T.H. Koornwinder and N.M. Muller,
Quantum double of a (locally) compact group,
Jour. of Lie theory 7, 33-35 (1997). Erratum, 187 (1998).
F.A. Bais, N.M. Muller and B.J. Schroers,
Quantum group symmetry and particle scattering in (2+1)-dimensional quantum gravity,
Nucl. Phys. B640, 3-45 (2002).


\bibitem{Comb}
A.Y. Alekseev, H. Grosse and V. Schomerus,
combinatorial quantisation of the Hamiltonian Chern-Simons theory I,
Comm. Math. Phys. 172, 317-458 (1995);
B.J. Schroers,
Cominatorial quantisation of Euclidean gravity in three dimensions,
Progress in Mathematics 198, 307-328 (2001).
C.~Meusburger and B.~J.~Schroers,
  Adv.\ Theor.\ Math.\ Phys.\  {\bf 7}, 1003 (2004)
  [arXiv:hep-th/0310218];
C.~Meusburger and B.~J.~Schroers,
  Class.\ Quant.\ Grav.\  {\bf 20}, 2193 (2003)
  [arXiv:gr-qc/0301108].




\bibitem{Witten}
E. Witten,
2+1 dimensional gravity as an exactly soluble problem,
Nucl. Phys. B311, 46-78 (1988).

\bibitem{Matschull}
H.J. Matschull,
On the relation between (2+1) Einstein gravity and Chern-Simons theory,
Class. Quant. Grav. 16, 2599-2609 (1999).

\bibitem{Ale}
A. Perez,
On the regularization ambiguities in loop quantum gravity,
Phys. Rev. D73, 044007 (2006).

\bibitem{schwartz}
A. Kirillov. Elements of the theory of representations. Translated from the Russian by Edwin Hewitt. Grundlehren der Mathematischen Wissenschaften,
220. Springer-Verlag, Berlin-New York, 1976.

\bibitem{FL2}
L. Freidel, E. Livine, unpublished.

\bibitem{Edmonds}
A.R. Edmonds,
Angular momentum in quantum mechanics,
Princeton Univ. Press (1957).

\bibitem{Noui2}
K. Noui,
A model for the motion of a particle in a quantum background,
preprint.

\bibitem{Japanese}
  H.~Komaie-Moghaddam, M.~Khorrami and A.~H.~Fatollahi,
  Loop diagrams in space with SU(2) fuzziness,
  arXiv:0712.2216 [hep-th];
S.~Imai and N.~Sasakura,
  Scalar field theories in a Lorentz-invariant three-dimensional
  noncommutative space-time,
  JHEP {\bf 0009}, 032 (2000);
Y.~Sasai and N.~Sasakura,
  Domain wall solitons and Hopf algebraic translational symmetries in
  noncommutative field theories,
  arXiv:0711.3059 [hep-th].

\bibitem{FM}
  L.~Freidel and S.~Majid,
  Noncommutative Harmonic Analysis, Sampling Theory and the Duflo Map in 2+1
  Quantum Gravity,
  arXiv:hep-th/0601004.

\bibitem{BS}
 B.J. Schroers,
 Lesson from (2+1)-dimensional quantum gravity,
 [arXiv:0710.5844].

\bibitem{bied}
L.C. Biedenharn, J.D. Louck, Angular momentum in quantum physics, Addison-Wesley
Publishing Company (1981).

\end{thebibliography}

\end{document}